\documentclass[aps,pre,twocolumn,groupedaddress,showpacs,preprintnumbers,amsmath,amssymb]{revtex4}

\usepackage{amsfonts}
\usepackage{amssymb}
\usepackage{psfrag}
\usepackage{float}
\usepackage{natbib}
\usepackage{amsmath}
\usepackage{float}
\usepackage{dcolumn}
\usepackage{lscape}
\usepackage{color}
\usepackage{stmaryrd}
\usepackage{epsfig}
\usepackage[dvips]{hyperref}

\def\be{\begin{equation}}
\def\ee{\end{equation}}
\def\bea{\begin{eqnarray}}
\def\eea{\end{eqnarray}}
\def\eq#1{(\ref{#1})}

\def\ms{\medskip}
\def\fig#1{Fig.\ \ref{#1}}
\def\figs#1{Figs.\ \ref{#1}}
\def\tab#1{Tab.\ \ref{#1}}
\def\sec#1{Sec.\ \ref{#1}}

\def\bfa{{\bf a}}
\def\bfr{{\bf r}}
\def\bfp{{\bf p}}

\def\d{{\rm d}}

\def\papa#1#2{\frac{\partial#1}{\partial#2}}
\def\Ai{{\rm Ai}}
\def\Cosh{{\rm cosh}}
\def\Sinh{{\rm sinh}}
\def\Tau{{\cal T}}
\def\simg{\,\hbox{\kern.1em \lower.6ex \hbox{$\sim$} \kern-1.12em
          \raise.6ex \hbox{$>$} }}
\def\siml{\,\hbox{\kern.1em \lower.6ex \hbox{$\sim$} \kern-1.12em
          \raise.6ex \hbox{$<$} }}
\def\lambdab{\widetilde{\lambda}}
\def\gb{\widetilde{g}}
\def\NPO{\rm{\text{NPO}}}

\newcommand{\Gft}{\widehat{G}}

\begin{document}

\title{Semiclassical theory
       for spatial density oscillations in fermionic systems}

\author{J. Roccia, M. Brack, and A. Koch}

\affiliation{Institute for Theoretical Physics, University of
             Regensburg, D-93040 Regensburg, Germany}


\begin{abstract}
We investigate the particle and kinetic-energy densities for a system
of $N$ fermions bound in a local (mean-field) potential $V(\bfr)$.
We generalize a recently developed semiclassical theory
[J. Roccia and M. Brack, Phys.\ Rev.\ Lett.\ {\bf 100}, 200408 (2008)],
in which the densities are calculated in terms of the closed orbits
of the corresponding classical system, to $D>1$ dimensions.
We regularize the semiclassical results $(i)$ for the U(1) symmetry
breaking occurring for spherical systems at $r=0$ and $(ii)$ near
the classical turning points where the Friedel oscillations are
predominant and well reproduced by the shortest orbit going
from $r$ to the closest turning point and back. 
For systems with spherical symmetry, we show that there exist
two types of oscillations which can be attributed to radial and
non-radial orbits, respectively. The semiclassical theory is tested 
against exact quantum-mechanical calculations for a variety of model 
potentials. We find a very good overall numerical agreement between 
semiclassical and exact numerical densities even for moderate particle 
numbers $N$. Using a ``local virial theorem'', shown to be valid 
(except for a small region around the classical turning points) for 
arbitrary local potentials, we can prove that the Thomas-Fermi 
functional $\tau_{\text{TF}}[\rho]$ reproduces the oscillations in 
the quantum-mechanical densities to first order in the oscillating 
parts.

\end{abstract}







\maketitle

\section{Introduction}
\label{secint}

Recent experimental success confining fermion gases in magnetic
traps \cite{jin} has led to renewed interest in theoretical studies
of confined degenerate fermion systems at zero
\cite{vig1,glei,bvz,mar1,mar2,mar3,vig2,homa,bm,mue} and finite
temperatures \cite{akde,zbsb}. According to the density functional
theory (DFT) \cite{hk,ks,dft}, the local particle density $\rho({\bfr})$
is the key ingredient of a system of interacting fermions in that it
contains all information about its ground state. In this paper we
study the oscillations in the particle density $\rho({\bfr})$
and in different forms of the kinetic-energy density of $N$ fermions
bound in a local potential $V(\bfr)$. Although we treat the particles
as non-interacting, we keep in mind that this potential models the
self-consistent Kohn-Sham (KS) potential obtained for an {\it interacting
system in the mean-field approximation}. 
We shall also consider
poten\-tials with infinitely steep walls, so-called ``billiards'',
which have been shown to be good approximations to the
self-consisten mean fields of quantum dots \cite{qdot} or metal
clusters \cite{sciam} with many particles.

A semiclassical theory for spatial density oscillations has been
developed recently in \cite{rb}. Using Gutz\-willer's semiclassical
Green function \cite{gubu}, expressions for the oscillating parts of
spatial densities of fermionic systems were given in terms
of the {\it closed orbits} of the corresponding classical system.
The semiclassical theory was shown in \cite{rb} to reproduce
very accurately the quantum oscillations in the spatial densities of
one-dimensional systems, even for moderate particle numbers $N$, and
some general results have also been given for arbitrary
higher-dimensional spherical potentials $V(r)$.

In this paper, we present in more detail the semiclassical
closed-orbit theory developed in \cite{rb} and apply it explicitly
for a variety of potentials in $D>1$ dimensions. We find overall
a good agreement between the quantum-mechanical and the
semiclassical densities.

The paper is organized as follows. In \sec{secbas} we give
the basic definitions of the quantum-mechanical spatial densities.
In \sec{secexa} we discuss the asymptotic (extended) Thomas-Fermi
(TF) limits for $N\to\infty$ and emphasize the existence of two
types of density oscillations occurring in potentials for $D>1$
with spherical symmetry (except for isotropic harmonic oscillators).

\sec{secscl} is devoted to the semiclassical closed-orbit theory
for spatial density oscillations. In Secs.\ \ref{secpot} -
\ref{secscl1dim} we review the basic equations and former results,
including also details that were not presented in \cite{rb}.
In Secs.\ \ref{secrad} and \ref{secnonrad} we extend the
semiclassical theory to higher-dimensional systems ($D>1$) and 
test its results for various model potentials against exact 
quantum-mechanical densities. 
In Sect.\ \ref{secregul} we discuss the regularization necessary in 
spherical systems for $D>1$ near the center ($r=0$), where a U(1) 
symmetry breaking occurs for $r>0$.
In a separate publication \cite{circ}, we have presented the analytical
determination and classification of all closed orbits in the
two-dimensional circular billiard and give analytical results of the
semiclassical theory for the spatial density oscillations in this
system. Some of the numerical results for the densities are included
in \sec{secdisc} of the present paper.

In \sec{secsurf} we present regularizations of the spatial densities 
near the classical turning points, where the semiclassical theory 
diverges, both for smooth potentials and for billiard systems.

\sec{traps} contains some general results valid for
finite fermion systems such as trapped fermionic gases or
metallic clusters. We discuss there, in particular, a ``local
virial theorem'' and, as its direct consequence, the extended
validity of the TF functional $\tau_{\text{TF}}[\rho]$.

Throughout this paper, we only treat the zero-tem\-perature ground state
of an $N$-particle system. In the Appendix \ref{appcor}, we outline how to
include finite temperatures for grand-canonical ensembles in the
semiclassical theory.

\vfill

\newpage

\section{The quantum-mechanical densities}
\label{secbas}

\subsection{Basic definitions and ingredients}
\label{secdef}

Let us recall some basic quantum-mechanical definitions, using the same
notation as in \cite{bm}. We start from the stationary Schr\"odinger
equation for particles with mass $m$, bound by a local potential
$V(\bfr)$ with a discrete energy spectrum $\{E_n\}$:
\be
\left\{-\frac{\hbar^2}{2m} \nabla^2 + V(\bfr)\right\} \phi_n(\bfr)
= E_n\, \phi_n(\bfr)\,.
\label{seq}
\ee
We order the spectrum and choose the energy
scale such that $0 < E_1 \leq E_2 \leq \dots \leq E_n \leq \dots$. We
consider a system with an even number $N$ of fermions with spin $s=1/2$
filling the lowest levels, and define the particle density by
\be
\rho(\bfr) \; := \; 2\!\! \sum_{E_n\leq \lambda}\!\! |\phi_n(\bfr)|^2,
\qquad \int \rho(\bfr)\,\d^Dr = N\,.
\label{rho}
\ee
Hereby $\lambda$ is the Fermi energy and the factor 2 accounts for the
fact that due to spin and time-reversal symmetry, each state $n$ is at
least two-fold degenerate. Further degeneracies, which may arise for
$D>1$, will not be spelled out but included in the summations over $n$.
For the kinetic-energy density, we consider two different definitions \cite{foot} 
\begin{eqnarray}
\tau(\bfr)   \; &:=& \; - \frac{\hbar^2}{2m}\; 2\!\! \sum_{E_n\leq \lambda}
                     \!\! \phi_n^*(\bfr)\nabla^2 \phi_n(\bfr)\,,
\label{tau}          \qquad\\
\tau_1(\bfr) \; &:=& \; \frac{\hbar^2}{2m}\; 2\!\! \sum_{E_n\leq \lambda}
                     \!\! |\nabla\phi_n(\bfr)|^2,
\label{tau1}
\end{eqnarray}
which upon integration both yield the exact total kinetic energy.
Due to the assumed time-reversal symmetry, the two above
functions are related by
\be
\tau(\bfr) = \tau_1(\bfr) - \frac12\, \frac{\hbar^2}{2m}\,
             \nabla^2\rho(\bfr)\,.
\label{taurel}
\ee
An interesting, and for the following discussion convenient quantity is their average
\be
\xi(\bfr) \; := \; \frac12\, [\tau(\bfr)+\tau_1(\bfr)]\,.
\label{xi}
\ee
For harmonic oscillators it has been observed \cite{bvz,bm,rkb1} that
inside the system (i.e., sufficiently far from the surface region),
$\xi(\bfr)$ is a smooth function of the coordinates, whereas $\tau(\bfr)$
and $\tau_1(\bfr)$, like the density $\rho(\bfr)$, exhibit characteristic
shell oscillations that are opposite in phase for $\tau$ and $\tau_1$.
We can express $\tau(\bfr)$ and $\tau_1(\bfr)$ in terms of $\xi(\bfr)$
and $\nabla^2\rho(\bfr)$:
\bea
\tau(\bfr) &=& \xi(\bfr) -\frac14\, \frac{\hbar^2}{2m}\,\nabla^2\rho(\bfr)\,,\label{tauxi2}\\
\tau_1(\bfr) &=& \xi(\bfr) +\frac14\, \frac{\hbar^2}{2m}\,\nabla^2\rho(\bfr)\,,
\label{tauxi}
\eea
so that $\rho(\bfr)$ and $\xi(\bfr)$ can be considered as the basic
densities characterizing our systems. Eqs.~\eq{rho} -- \eq{tauxi} are
exact for arbitrary potentials $V\bf(r)$. For any even number $N$ of
particles they can be computed once the quantum-mechanical wave functions
$\phi_n(\bfr)$ are known. As mentioned in the introduction, the potential
$V(\bfr)$ can be considered to represent the self-consistent mean field
of an interacting system of fermions obtained in the DFT approach.
The single-particle wavefunctions $\phi_n(\bfr)$ are then the Kohn-Sham
orbitals \cite{ks} and $\rho(\bfr)$ is (ideally) the ground-state particle
density of the interacting system.

For later reference we express the densities \eq{rho}--\eq{tau1} in
terms of the Green function in the energy representation, which in the
basis $\left\{\phi_n(\bfr)\right\}$ is given by
\be
G(E,\bfr,\bfr')= \sum_n \frac{\phi_n^{\star}(\bfr)
                 \phi_n(\bfr')}{E+i \epsilon -E_n} \,,
                 \qquad (\epsilon >0) \,.
\label{green}
\ee
Using the identity $\displaystyle 1/(E+i \epsilon -E_n)={\cal P}
[1/(E-E_n)]-i \pi \delta (E-E_n)$, where ${\cal P}$ is the Cauchy
principal value, one can write the densities as
\begin{eqnarray}
\rho({\bf r})&=&-\frac{1}{\pi}\, \text{Im} \int_0^{\lambda} \d E \, G(E,{\bf r},{\bf r}')
|_{{\bf r'}={\bf r}}
\,,  \label{rhog} \\
\tau({\bf r})&=& \frac{\hbar^2}{2\pi m}\, \text{Im} \int_0^{\lambda}  \d E \,
\nabla^2_{{\bf r}'}G(E,{\bf r},{\bf r}') |_{{\bf r}'={\bf r}}
 \,, \label{taug}\\
\tau_1({\bf r})&=&
-\frac{\hbar^2}{2\pi m}\, \text{Im} \int_0^{\lambda} \d E  \,
\nabla_{{\bf r}}\nabla_{{\bf r}'}G(E,{\bf r},{\bf r}')  |_{{\bf r}'={\bf r}}
 \,, \ \ \ \ \label{tau1g}
\end{eqnarray}
whereby the subscript of the nabla operator $\nabla$ denotes the
variable on which it acts.

The density of states $g(E)$ of the system \eq{seq} is given by a sum
of Dirac delta functions, which can be expressed as a trace integral
of the Green function:
\be
g(E) = \sum_n \delta(E-E_n) = -\frac{1}{\pi}\, \text{Im} \int \d^D r
                              \,G(E,\bfr,\bfr')|_{{\bf r}'={\bf r}}\,.
\label{dos}
\ee
The particle number can then also be obtained as
\be
N = N(\lambda) = 2\int_0^\lambda \d E\,g(E)\,.
\label{intdos}
\ee
Due to the discreteness of the spectrum, $N(\lambda)$ is a monotonously
increasing staircase-function and consequently the function $\lambda(N)$,
too, is a monotonously increasing staircase-function.

\subsection{Asymptotic quantum-mechanical results}
\label{secexa}

\subsubsection{Thomas-Fermi limits and oscillating parts}
\label{sectf}

In the limit $N\to\infty$, the densities are expected to go over
into the approximations obtained in the Thomas-Fermi (TF) theory
\cite{matf}. These are given, for any local potential $V(\bfr)$, by
\be
\rho_{\text{TF}}(\bfr)  =  \frac{4}{D}\,\frac{1}{\Gamma(\frac{D}{2})}
                   \left(\frac{m}{2\pi\hbar^2}\right)^{\!D/2}
                   [\lambda_{\text{TF}}-V(\bfr)]^{D/2}\,,
\label{rhotf}
\ee
\vspace{-.5cm}
\be
\hspace{-3.5cm}(\tau_1)_{\text{TF}}(\bfr) =  \xi_{\text{TF}}(\bfr)
                \; = \; \tau_{\text{TF}}(\bfr) \,,
\label{xitf}
\ee
\vspace{-.6cm}
\be
\tau_{\text{TF}}(\bfr)   =  \frac{4}{(D\!+\!2)}\,\frac{1}{\Gamma(\frac{D}{2})}
                   \left(\frac{m}{2\pi\hbar^2}\right)^{\!D/2}
                   \![\lambda_{\text{TF}}-V(\bfr)]^{D/2+1}\,.
\label{tautf}
\ee
These densities are defined only in the classically allowed regions
where $\lambda_{\text{TF}}\geq V(\bfr)$, and the Fermi energy $\lambda_{\text{TF}}$ is
defined such as to yield the correct particle number $N$ upon integration
of $\rho_{\text{TF}}(\bfr)$ over all space. The direct proof that the
quantum-mechanical densities, as defined in \sec{secbas}
in terms of the wavefunctions of a smooth potential, reach the above TF
limits for $N\to\infty$ is by no means trivial. It has been given for
isotropic harmonic oscillators in arbitrary dimensions in Ref.\ \cite{bm}.

The TF densities \eq{rhotf}--\eq{tautf} fulfill the following
functional relation:
\begin{eqnarray}
\hspace{-.4cm}
\tau_{\text{TF}}(\bfr) &=& \tau_{\text{TF}}[\rho_{\text{TF}}(\bfr)]\,\nonumber\\
&=& \frac{\hbar^2}{2m} \frac{4\pi D}{(D+2)}\!\left[\frac{D}{4}
    \Gamma\left(\frac{D}{2}\right)\right]^{\!2/D}\!\rho_{\text{TF}}^{1+2/D}(\bfr)\,,~~~
\label{tautff}
\end{eqnarray}
which will be investigated further below.

For smooth potentials in $D>1$ dimensions, next-to-leading order terms
in $1/N$ modify the smooth parts of the spatial densities, which are
obtained in the extended Thomas-Fermi (ETF) model as corrections of
higher order in $\hbar$ through an expansion in terms of gradients
of the potential \cite{kirk}. These corrections usually
diverge at the classical turning points and can only be used
sufficiently far from the turning points, i.e., in the interior of the
system. We do not reproduce the ETF densities here but refer to
\cite{book} (chapter 4) where they are given for arbitrary smooth
potentials in $D=2$ and 3 dimensions, and to \cite{bm} where explicit
results are given for spherical harmonic oscillators in $D=2$ and 4
dimensions.

This leads us to decompose the densities in the following way:
\bea
\rho(\bfr)   & = & \rho_{\text{(E)TF}}(\bfr) + \delta\rho(\bfr)\,,
\label{rhodec}\\
\tau(\bfr)   & = & \tau_{\text{(E)TF}}(\bfr) + \delta\tau(\bfr)\,,
\label{taudec}\\
\tau_1(\bfr) & = & (\tau_1)_{\text{(E)TF}}(\bfr) + \delta\tau_1(\bfr)\,,
\label{tau1dec}\\
\xi(\bfr)    & = & \xi_{\text{(E)TF}}(\bfr) + \delta\xi(\bfr)\,.
\label{xidec}
\eea
For $D=1$ and for billiard systems \cite{notebil}, the subscripts
TF and the explicit relations \eq{rhotf} -- \eq{tautf} hold. The
oscillating parts $\delta\rho(\bfr)$ etc.\ are the main objects of
this paper.

\subsubsection{Two types of oscillating parts in spherical systems}
\label{secsep}

\begin{figure}[h]
\includegraphics[width=1.15\columnwidth,clip=true]{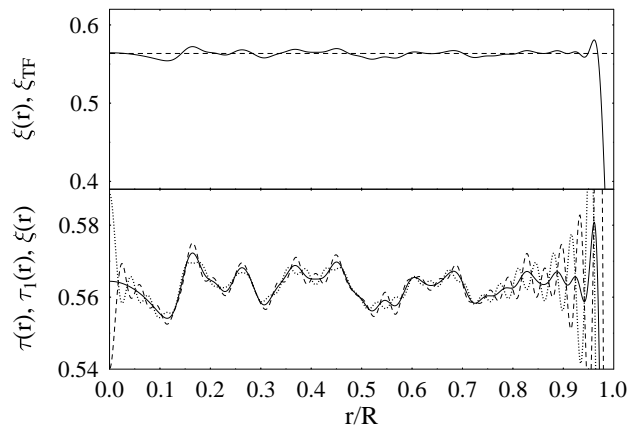}\vspace{-0.3cm}
\caption{\label{osc}
Kinetic-energy density profiles of a $3D$ spherical billiard with $N=100068$
particles (units: $\hbar^2\!/2m=R=1$; all densities are divided by
$N^{5/3}$). {\it Upper panel:} $\xi(r)$ (solid
line) and its constant TF value $\xi_{\text{TF}}$ (dashed).
{\it Lower panel:} $\tau(r)$ (dashed), $\tau_1(r)$ (dotted) and
$\xi(r)$ (solid line). Note that in both panels, the vertical
scale does not start at zero.
}
\end{figure}

We have investigated the density oscillations in various potentials
in $D>1$ dimensions with {\it radial symmetry} such that
$V(\bfr)=V(r)$, where $r=|\bfr|$. We found that, generally, there exist
two types of oscillations in their spatial densities:\\ ($i$) regular,
short-ranged oscillations with a constant wavelength in the radial
variable $r$ over the whole region, and\\ ($ii$) irregular, long-ranged
oscillations whose wavelength decreases with increasing $r$.\\
An example is shown in \fig{osc} for a spherical billiard with unit
radius containing $N=100068$ particles. Note the irregular,
long-ranged oscillations of $\xi(r)$ around its bulk value \cite{notebil}
$\xi_{\text{TF}}$ seen in the upper panel. In the lower
panel, where we exhibit only an enlarged region around the bulk
value, we see that $\tau(r)$ and $\tau_1(r)$ oscillate regularly
around $\xi(r)$, but much faster than $\xi(r)$ itself and with
opposite phases. The same two types of oscillations are also
found in the particle density $\rho(r)$.

For radial systems, we can thus decompose the oscillating parts of the
spatial densities defined in \eq{rhodec} -- \eq{xidec} as follows:
\bea
\delta\rho(r)   & = & \delta_{\text{r}}\rho(r) + \delta_{\text{irr}}\rho(r)\,,
\label{drhodec}\\
\delta\tau(r)   & = & \delta_{\text{r}}\tau(r) + \delta_{\text{irr}}\tau(r)\,,
\label{dtaudec}\\
\delta\tau_1(r) & = & \delta_{\text{r}}\tau_1(r) + \delta_{\text{irr}}\tau_1(r)\,,
\label{dtau1dec}\\
\delta\xi(r)    & = & \delta_{\text{irr}}\xi(r)\,.
\label{dxidec}
\eea
Here the subscript ``r'' denotes the regular, short-ranged parts
of the oscillations, while their long-ranged, irregular parts are denoted
by the subscript ``irr''. We emphasize that this separation of the
oscillating parts does not hold close to the classical turning points.

As we see in \fig{osc} and in later examples, the oscillating parts
defined above fulfill the following properties in the interior of
the system (i.e., except for a small region around the classical
turning points):

\ms

\noindent
a) For $D>1$, the irregular oscillating parts of $\tau(r)$ and $\tau_1(r)$
are asymptotically identical and equal to $\delta\xi(r)$:
\be
\delta_{\text{irr}}\tau(r) \;\simeq\; \delta_{\text{irr}}\tau_1(r)
              \;\simeq\; \delta_{\text{irr}}\xi(r) = \delta\xi(r)\,.
\ee

\noindent
b) The irregular oscillations are {\it absent} (i.e., asymptotically zero)
in the densities of all potentials in $D=1$ and, also, in isotropic
harmonic oscillators (see Ref.\ \cite{bm}) and in linear potentials (see
\cite{nlvt}) for arbitrary $D$.

\ms

\noindent
c) The regular oscillating parts of $\tau(r)$ and $\tau_1(r)$ are
asymptotically equal with opposite sign:
\be
\delta_{\text{r}}\tau(r) \;\simeq\; -\, \delta_{\text{r}}\tau_1(r)\,.
\ee
(This relation holds in particular for isotropic harmonic oscillators, for
which it has been derived \cite{bm} asymptotically for $N\to\infty$
from quantum mechanics.)

These numerical findings will be understood and explained
within the semiclassical theory developed in the following.

Henceforth, the symbol $\delta$ will always denote the
sum of both types of oscillating parts and the subscripts will
only be used if reference is made to one particular type of
oscillations.


\section{Semiclassical closed-orbit theory}
\label{secscl}

In this section we present the semiclassical theory, initiated by
Gutzwiller (see \cite{gutz} and earlier references quoted therein,
and \cite{gubu}), for the approximate
description of quantum oscillations in terms of classical orbits.
In \sec{secpot} we recall the trace formula for the density of
states, and in \sec{secsclden} we present the newly developed theory
for spatial density oscillations \cite{rb}. In both cases, we limit
ourselves -- as in the previous section -- to $N$ non-interacting
fermions in a local potential $V(\bfr)$. The inclusion of finite
temperatures in the semiclassical theory is dealt with in Appendix
\ref{appcor}.

\subsection{Brief review of periodic orbit theory for the density of states}
\label{secpot}

Before deriving semiclassical expressions for the spatial densities,
we remind the reader of the periodic orbit theory (POT) for the
density of states. The starting point is the semiclassical approximation
of the Green function \eq{green} which was derived by Gutzwiller
\cite{gubu}:
\begin{equation}
G_{\text{scl}}(E,{\bf r,r'}) = \alpha_{_D}\sum_{\gamma} \sqrt{|{\cal D}_\gamma|}
                               \,e^{\frac{i}{\hbar}
                               S_\gamma(E,{\bf r,r'})-i \mu_\gamma\frac{\pi}{2}}.
\label{sclgreen}
\end{equation}
The sum runs over all classical trajectories $\gamma$ leading from
a point $\bfr$ to the point $\bfr'$ at fixed energy $E$.
$S_\gamma(E,{\bf r,r'})$ is the action integral taken along the
trajectory $\gamma$
\be
S_\gamma(E,{\bf r,r'}) = \int_{\bfr}^{\bfr'} {\bf p}(E,{\bf q})\cdot \d \,{\bf q}\,,
\label{actint}
\ee
whereby ${\bf p}(E,{\bf r})$ is the classical momentum
\be
{\bf p}(E,{\bf r}) = \frac{\dot{{\bf r}}}{|{\dot{\bf r}}|}
                     \sqrt{2m[E-V(\bfr)]}\,,
\label{pclass}
\ee
defined only inside the classically allowed region where $E\ge V(\bfr)$;
its modulus is denoted by $p(E,\bfr)$.
${\cal D}_\gamma$ is the Van Vleck determinant:
\be
{\cal D}_\gamma = \frac{(-1)^D\,m^2}{p(E,{\bf r})\,p(E,{\bf r'})}\,{\cal D}_{\bot}\,,
                  \qquad
{\cal D}_{\bot} = \det (\partial{\bf p}_{\bot}/\partial{\bfr'}_{\!\bot})\,,
\label{vleckdet}
\ee
where $\bfp_{\bot}$ and $\bfr_{\!\bot}'$ are the initial momentum and
final coordinate, respectively, {\it transverse} to the orbit $\gamma$.
The Morse index $\mu_\gamma$ counts the sign changes of the
eigenvalues of the Van Vleck determinant along the trajectory $\gamma$
between the points $\bfr$ and $\bfr'$; it is equal to the number of
conjugate points along the trajectory \cite{conjp}.
The prefactor in
\eq{sclgreen} is given by
\be
\alpha_D=2 \pi (2 i \pi \hbar)^{-(D+1)/2}.
\ee

The approximation \eq{sclgreen} of the Green function is now inserted
into the r.h.s of \eq{dos} for the density of states $g(E)$.
Since $\bfr'=\bfr$ in the trace integral of \eq{dos}, only closed orbits
contribute to it. The running time $T_\gamma(E,\bfr)$ of these orbits,
i.e., the time it takes the classical particle to run
though the closed orbit, is given by
\be
T_\gamma(E,\bfr) = \frac{\d S_\gamma(E,\bfr,\bfr)}{\d E}\,.
\label{time}
\ee
It was shown by Berry and Mount \cite{bemo} that to leading order in
$\hbar$, the orbits with zero running time, $T_\gamma(E,\bfr)=0$, yield the
smooth TF value of $g(E)$. In systems with $D>1$ higher-order terms in
$\hbar$ also contribute, which can also be obtained from the ETF model
(see, e.g., chapter 4 of \cite{book}). Separating smooth and
oscillatory parts of the density of states by defining
\be
g(E):=\gb(E)+\delta g(E)\,,
\label{dossep}
\ee
the oscillating part $\delta g(E)$ is, to leading order in $\hbar$,
given by the {\it semiclassical trace formula}
\be
\delta g(E) \simeq \sum_{\text{PO}} {\cal A}_{\text{PO}}(E) \cos\left[
                   \frac{1}{\hbar}\,S_{\text{PO}}(E)-\frac{\pi}{2}\,\sigma_{\text{PO}}\right],
\label{trf}
\ee
where the sum runs over all {\it periodic orbits} (POs). For systems in
which all orbits are isolated in phase space, Gutzwiller \cite{gutz}
derived explicit expressions for the amplitudes ${\cal A}_{\text{PO}}(E)$,
which depend on the stability of the orbits, and for the Maslov
indices $\sigma_{\text{PO}}$. Performing the trace integral in \eq{dos}
along all directions transverse to each orbit $\gamma$ in the stationary
phase approximation (SPA) leads immediately to the periodicity of the
contributing orbits. The Maslov index $\sigma_{\text{PO}}$ collects all
phases occurring in \eq{sclgreen} and in the SPA for the trace integral
(see \cite{masl} for detailed computations of $\sigma_{\text{PO}}$).
It has been shown \cite{crl} that $\sigma_{\text{PO}}$ is a
canonical and topological invariant property of any PO.
$S_{\text{PO}}(E)$ is the closed action integral
\be
S_{\text{PO}}(E) = \oint_{\text{PO}} \bfp(E,{\bf q})\cdot \d\, {\bf q}\,.
\label{spo}
\ee
For smooth one-dimensional potentials, the trace formula is
particularly simple and reads
\be
\delta g^{(D=1)}(E) = \frac{T_1(E)}{\pi\hbar}\,\sum_{k=1}^\infty
                      (-1)^k \cos\left[\frac{k}{\hbar}\,S_1(E)\right],
\label{trf1}
\ee
where the sum is over the repetitions $k\geq 1$ of the primitive
orbit with action $S_1(E)$ and period $T_1(E)=S'_1(E)$. Equation
\eq{trf1} is equivalent to the sum of delta functions in \eq{dos},
using the spectrum obtained in the WKB approximation \cite{book,beta}.
For systems with $D>1$ with continuous symmetries (and hence also
for integrable systems), the same type of trace formula \eq{trf}
holds, but the summation includes all degenerate families of
periodic orbits and the amplitudes ${\cal A}_{\text{PO}}(E)$ and indices
$\sigma_{\text{PO}}$ have different forms. For an overview of various
trace formulae and the pertinent literature, as well as many
applications of the POT, we refer to \cite{book}.

\subsection{Semiclassical approximation to the spatial densities}
\label{secsclden}

In order to derive semiclassical expressions for the spatial
densities defined in \sec{secbas}, we start from the
expressions given in the equations \eq{rhog} -- \eq{tau1g},
which are functions of $\bfr$ and the Fermi energy $\lambda$,
and replace the exact Green function $G(E,{\bf r,r'})$ by
its semiclassical expansion \eq{sclgreen}. The energy
integration can be done by parts, using \eq{vleckdet} and
\eq{time}, and to leading order in $\hbar$ we obtain for the
particle density
\be
\rho(\lambda,{\bf r}) \simeq \frac{2m\hbar}{\pi\,p(\lambda,\bfr)}\, \text{Re} \ \alpha_{_D}\!\sum_\gamma\!
   \frac{\sqrt{|{\cal D}_{\bot}|}_{{\bf r'}={\bf r}}}{T_\gamma(\lambda,{\bf r})}
   \,e^{\frac{i}{\hbar} S_\gamma(\lambda,{\bf r,r})-i \mu_\gamma\frac{\pi}{2}}.
\label{rhosc}
\ee
Again, the orbits with zero running time $T(E,\bfr)=0$ yield, to leading
order in $\hbar$, the smooth TF particle density \eq{rhotf}; the proof given
in \cite{bemo} for the density of states applies also to the spatial
densities discussed here. Like for the density of states, higher-order
$\hbar$ corrections contribute also to the smooth part of
$\rho(\bfr)$ in $D>1$ and will be included in their ETF expressions.
The periodic orbits (POs), too, can only contribute to the
smooth part of $\rho(\bfr)$, since their action integrals \eq{spo}
are independent of $\bfr$ and hence the phase in the exponent of
\eq{rhosc} is constant. Thus, {\it a priori} only {\it non-periodic
orbits} (NPOs)
contribute to the oscillating part of $\rho(\bfr)$. The same holds
also for the other spatial densities, so that we can write their
semiclassical approximations as \cite{rb}:
\bea
&&\hspace{-1.2cm}
\delta\rho(\bfr)  \simeq
                  \frac{2m\hbar}{\pi\,p(\lambdab,{\bf r})}\, \text{Re} \ \alpha_{_D}\sum_{\NPO}\;
                  \frac{\sqrt{|{\cal D}_{\bot}|}_{{\bf r'}={\bf r}}}{T(\lambdab,{\bf r})}
                  \,e^{\Phi(\lambdab,{\bf r})}\,,
\label{drhosc}\\
&&\hspace{-1.2cm}
\delta \tau({\bf r}) \simeq \frac{\hbar\,p(\lambdab,{\bf r})}{\pi}\, \text{Re} \
                     \alpha_{_D}\sum_{\rm NPO}\;
                     \frac{\sqrt{|{\cal D}_{\bot}|}_{{\bf r}'={\bf r}}}{T(\lambdab,{\bf r})}\,
                     e^{i\Phi(\lambdab,{\bf r})}\,,
\label{dtausc}
\eea
\vspace{-.6cm}
\be
\delta \tau_1({\bf r}) \simeq  \frac{\hbar\,p(\lambdab,\bfr)}{\pi}\,
                       \text{Re} \ \alpha_{_D}\sum_{\rm{\text{NPO}}} Q(\lambdab,\bfr)\,
                       \frac{\sqrt{|{\cal D}_{\bot}|}_{{\bf r}'={\bf r}}}
                       {T(\lambdab,{\bf r})}\, e^{i\Phi(\lambdab,{\bf r})}\,.
\label{dtau1sc}
\ee
The sums are only over {\it non-periodic orbits} (NPOs) that lead
from a point $\bfr$ back to the same point $\bfr$. For convenience,
we have omitted the subscript ``NPO'' from all quantities in the above
equations. The phase function $\Phi(\lambdab,\bfr)$ is given by
\be
\Phi(\lambdab,{\bf r})=S(\lambdab,{\bf r,r})/\hbar-\mu\frac{\pi}{2}\,.
\label{phase}
\ee
The quantity $Q(\lambdab,\bfr)$ appearing in \eq{dtau1sc} for
$\delta\tau_1(\bfr)$ is defined as
\be
Q(\lambdab,\bfr) = \frac{\left[{\bf p}(\lambdab,\bfr)
                   \cdot{\bf p}(\lambdab,\bfr')\right]_{\bfr'=\bfr}}
                   {p^2(\lambdab,\bfr)}
                 = \cos[\,\theta({\bfp,\bfp'})\,]\,,
\label{mismatch}
\ee
where $\bfp$ and $\bfp'$ are the short notations for the initial and
final momentum, respectively, of a given closed orbit $\gamma$ at the
point $\bfr$. These are obtained also from the action integral \eq{actint}
by the canonical relations
\be
\left. \nabla_{\bfr}S_\gamma(\lambdab,{\bf r,r'})\right|_{\bfr=\bfr'} = -\bfp\,,\quad
\left. \nabla_{\bfr'}S_\gamma(\lambdab,{\bf r,r'})\right|_{\bfr=\bfr'} = \bfp'\,.
\label{pcanon}
\ee
Since $Q$ in \eq{mismatch} depends on the angle $\theta$ between $\bfp$
and $\bfp'$, it may be called the ``momentum mismatch function'',
being +1 for $\bfp=\bfp'$ (i.e., for POs) and $-1$ for $\bfp=-\bfp'$
(e.g., for self-retracing NPOs).

Note that the upper limit $\lambda$ of the energy integral in
\eq{rhog} -- \eq{tau1g} has been replaced here by the smooth Fermi
energy $\lambdab$ defined by
\be
N = 2\int_0^{\lambdab} \d E\,\gb(E)\,,\qquad
\lambda = \lambdab + \delta\lambda\,.
\label{nsmooth}
\ee
The reason for this is the following. Since $\lambda(N)$ is a
non-smooth staircase function, as mentioned
at the end of \sec{secbas}, it is natural to expand it
around its smooth part $\lambdab$ which can be identified with its
TF value $\lambda_{\text{TF}}$ (or $\lambda_{\text{ETF}}$ for $D>1$).
Taylor expanding equation \eq{intdos} using \eq{dossep} up to first order
in $\delta\lambda$, we easily obtain an expression for its oscillating part
(cf.\ \cite{clmrs}):
\be
\delta\lambda \simeq -\frac{1}{g_{\text{ETF}}(\lambdab)}\int_0^{\lambdab} \d E\,
                     \delta g(E)\,.
\ee
The quantity $\delta\lambda$ is of higher order in $\hbar$ than $\lambdab$
and can be considered as a small semiclassical correction; the $\delta
g(E)$ in the integrand may be expressed through the trace formula \eq{trf}.
Now, the contribution of the zero-length orbits to \eq{rhosc} yields
formally the smooth (E)TF density, but taken at the exact (quantum) value
of $\lambda$. The density should therefore be developed around the smooth
(E)TF value $\lambdab$ before it can be identified with the standard
(E)TF density. Its first variation with $\delta\lambda$ leads to a
further smooth contribution which should be taken into account. The
same holds for the other densities. The contribution of all
finite-length orbits to \eq{rhosc} is of higher order in $\hbar$
than the leading smooth (ETF) terms, so it is consistent to evaluate
them at $\lambdab$.

In one-dimensional systems, all smooth terms can be exactly
controlled. The smooth part of the density may be written as
\be \left.
\rho_{\text{TF}}(\lambda,x) \simeq \rho_{\text{TF}}(\lambdab,x) + \delta\lambda\,
     \frac{\d\rho_{\text{TF}}(\lambda,x)}{\d\lambda}\right|_{\lambdab}.
\label{rhotftay}
\ee
The first term on the r.h.s. is the standard TF density (for $D=1$).
The second term, using \eq{trf1} and the fact that $g_{\text{TF}}(\lambda_{\text{TF}})
=T_1(\lambda_{\text{TF}})/2\pi\hbar$ for $D=1$, is found to exactly cancel
the contribution of the periodic orbits to \eq{rhosc} (evaluated at
$\lambdab$), which has been explicitly calculated in \cite{rb}
and given in Eq.\ (22) there.

For $D>1$ dimensions, we cannot prove that the same cancellation of
smooth terms takes place. Furthermore, for the circular billiard
treated in \cite{circ} it is shown that the contributions of periodic
and non-periodic orbits cannot be separated in the vicinity of
bifurcations that occur for $D>1$ under variation of $\bfr$. For
arbitrary local potentials in $D>1$ dimensions, it is in general
a difficult task to evaluate all nonperiodic closed orbits. In
non-integrable systems, the number of POs is known to grow
exponentially with energy or some other chaoticity parameter
(cf.\ the Appendix H in \cite{chaos2} or, to a large extent,
\cite{chaos3}); the number of NPOs is evidently even much larger.

For the semiclassical density of states \eq{trf}, the summation over POs
is known not to converge in general (cf.\ \cite{chaos4}). For the
semiclassical expressions \eq{drhosc} -- \eq{dtau1sc}, however, the
convergence of the sums over NPOs is appreciably improved due to the
appearance of their periods $T(\lambdab,\bfr)$ in the denominators. In
practice, we find that it is sufficient to include only a finite number
of shortest orbits, as illustrated for example \ in \fig{disk606} below.

The expressions \eq{drhosc} -- \eq{dtau1sc} are only valid
if the NPOs going through a given point $\bfr$ are {\it isolated}.
In systems with continuous symmetries, {\it caustic points} exist in
which the Van Vleck determinant ${\cal D}_\perp$ becomes singular.
The same happens at points where {\it bifurcations} of NPOs occur.
In such cases, {\it uniform approximations} can be developed which
lead to finite semiclassical expressions; these will be presented
in Sec. \ref{secregul} and in \cite{circ}.

We should also emphasize that the semiclassical approximations
are not valid in regions close to the classical turning points
$\bfr_\lambda$ defined by $V(\bfr_\lambda)=\lambdab$. Since the
classical momentum $p(\lambdab,\bfr_\lambda)$ in \eq{pclass}
becomes zero there, the spatial density \eq{drhosc} always
diverges at the turning points. Furthermore the running time
$T(\lambdab,{\bf r})$, which appears in the denominator of all
densities \eq{drhosc} -- \eq{dtau1sc}, may turn to zero at the
turning point for certain orbits. To remedy these divergences,
one has to resort to the technique of linearizing a smooth potential
$V(\bfr)$ around the classical turning points, which is familiar
from WKB theory \cite{wkb}. We shall discuss this in detail
in \sec{secsurf}.

Our semiclassical formulae \eq{drhosc} -- \eq{dtau1sc} can also
be applied to billiard systems in which a particle moves freely
inside a given domain and is ideally reflected at its boundary.
The only modification is that for a given orbit, each reflection
at the boundary contributes one extra unit to the Morse index
$\mu$ in \eq{phase}, since the difference in the semiclassical
reflection phases between a soft and a hard wall is $\pi/2$.
A detailed application of our formalism to the two-dimensional
circular billiard, including a complete determination of all
closed orbits of this system, has been given in \cite{circ}.

\subsection{Local virial theorem}
\label{secdlvt}

\subsubsection{Statement and test of the theorem}

We now shall discuss a result which can be directly inferred
from the semiclassical equations \eq{drhosc} -- \eq{dtau1sc},
without detailed knowledge of the NPOs that contribute to them
in a particular potential.

Since the modulus of the momentum $p(\lambdab,\bfr)$ depends only on
position and Fermi energy, but not on the orbits, we have taken it
outside the sum over the NPOs. Comparing the prefactors in \eq{drhosc}
and \eq{dtausc} and using \eq{pclass}, we immediately find \cite{rb}
the relation
\begin{equation}
\delta \tau({\bf r})\simeq[\lambdab-V({\bf r})] \,\delta \rho({\bf r}) \,.
\label{lvt}
\end{equation}
This is exactly the local virial theorem (LVT) that was derived in
\cite{bm} from the quantum-mechanical densities in the asymptotic
limit $N\to\infty$ for isotropic harmonic oscillators. Here we
obtain it explicitly from our semiclassical approximation. Since no
further assumption about the potential or the contributing NPOs has
been made, the LVT \eq{lvt} holds for arbitrary integrable or 
non-integrable systems in arbitrary dimensions with local potentials 
$V(\bfr)$ and hence also for {\it interacting fermions in the 
mean-field approximation given by the DFT}. We recall, however, that 
\eq{lvt} is not expected to be valid close to the classical turning 
points.

No such theorem holds for the density $\delta\tau_1({\bf r})$, since it
depends on the relative directions of the momenta $\bfp$ and $\bfp'$
of each contributing orbit through the factor $Q(\lambdab,\bfr)$
\eq{mismatch} appearing under the sum in \eq{dtau1sc}.
\begin{figure}[h]
\includegraphics[width=1\columnwidth,clip=true]{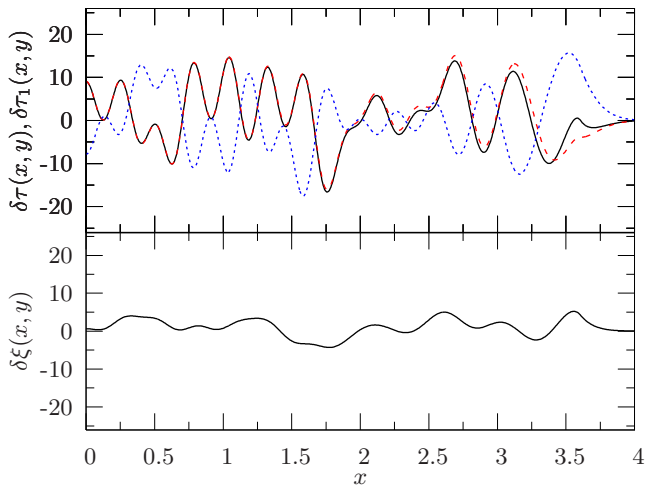}
\caption{\label{chaos}
(Color online) Oscillating part of spatial densities of $N=632$
particles in the nearly chaotic potential \eq{vqo} with $\kappa=0.6$
($\hbar=m=1$).
{\it Top:} The solid (black) line gives the r.h.s. of the LVT \eq{lvt},
the dashed (red) line gives $\delta \tau(x,y)$, and the dotted (blue)
line gives $\delta \tau_1(x,y)$, all taken along the line $y=x/\sqrt{3}$.
{\it Bottom:} $\delta \xi(x,y)$ along $y=x/\sqrt{3}$.
}
\end{figure}

In \fig{chaos} we test \eq{lvt} explicitly for the coupled two-dimensional
quartic oscillator
\be
V(x,y)=\frac{1}{2}(x^4+y^4)-\kappa\, x^2 y^2\,,
\label{vqo}
\ee
whose classical dynamics is almost chaotic in the limits $\kappa=1$ and
$\kappa\to -\infty$ \cite{btu,erda}, but in practice also for $\kappa=0.6$
(see, e.g., \cite{marta}). We have computed its wavefunctions using the
code developed in \cite{marta}.
In the upper panel of \fig{chaos} we show the left side (dashed line) and
the right side (solid line) of the LVT \eq{lvt} for this system with $N=632$
particles, using the exact densities along line $y=x/\sqrt{3}$, i.e.,
$\delta\rho(x,x/\sqrt{3})$ and $\delta\tau(x,x/\sqrt{3})$. The
agreement between both sides is seen to be very good, except in the
surface region. We also show $\delta\tau_1(x,x/\sqrt{3})$ (dotted line).
This demonstrates that the leading contributing NPOs in this system
are not self-retracing. Correspondingly, the quantity
$\delta\xi(x,x/\sqrt{3})$ in the lower panel is seen not to be negligible.


\subsection{$D=1$ dimensional systems}
\label{secscl1dim}

In a one-dimensional potential $V(x)$ there is only
linear motion along the $x$ axis. As discussed in \cite{rb},
the only types of NPOs are those running from a given point
$x$ to one of the turning points and back, including $k\geq 0$
full periodic oscillations between both turning points.
We name the two types of orbits the ``+'' orbits that start
from any point $x\neq 0$ towards the {\it closest} turning
point and return to $x$, and the ``$-$'' orbits that are
first reflected from the {\it farthest} turning point.
Clearly, these orbits have opposite initial and final momenta:
$p=-p'$, so that the momentum mismatch function \eq{mismatch}
equals $Q(\lambdab,x)=-1$. Consequently, one obtains from
\eq{dtau1sc} directly the relations
\be
\delta\tau_1(x) \simeq -\delta\tau(x)\,, \qquad \delta\xi(x) \simeq 0\,.
\label{tautau1}
\ee
Note that these results do not hold near the classical
turning points, where the semiclassical approximation breaks
down (cf.\ \sec{secsurf}; see also the example in \fig{x4friedun},
where $\delta\xi(x)$ is small inside the systems but becomes
comparable to $\delta\rho(x)$ near the turning points.)

The explicit evaluation of \eq{drhosc} for $D=1$ was done in
\cite{rb} for smooth potentials; the result in the present notation is
\be
\delta\rho(x) \simeq -\frac{m}{\pi p(\lambdab,x)}
                     \sum_{k=0, \pm}^\infty (-1)^k
                     \frac{\cos[k S_\pm^{(k)}(\lambdab,x)/\hbar]}
                          {T_\pm^{(k)}(\lambdab,x)}\,,
\label{drhosc1}
\ee
where $S_\pm^{(k)}$ are the actions of the ``+'' and ``$-$''
type NPOs (including $k$ full periods), and $T_\pm^{(k)}$
are there running times defined by \eq{time}.

\begin{figure}[h]
\includegraphics[width=0.95\columnwidth,clip=true]{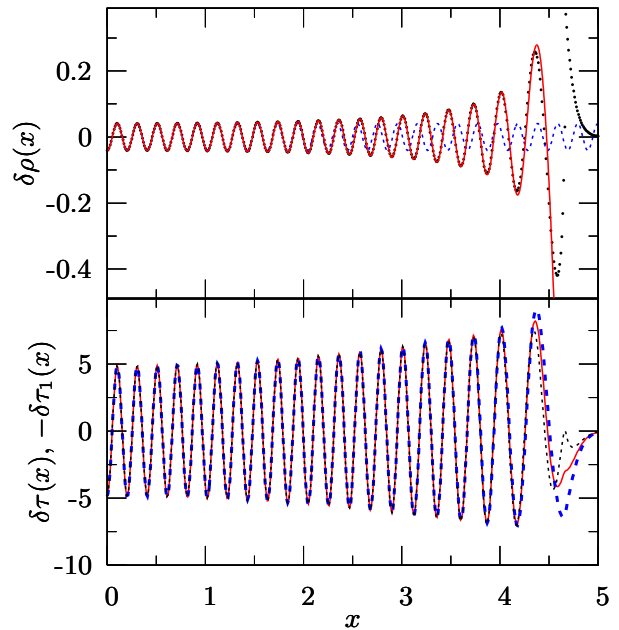}\vspace*{-0.2cm}
\caption{\label{1dx4drho}
(Color online) {\it Upper panel:} Oscillating part $\delta\rho(x)$ of
the particle density of $N$=40 particles in the quartic potential \eq{Vx4}
(without spin degeneracy; units: $\hbar=m=1$). Dots (black)
show the quantum-mechanical result; the solid line (red) shows the
semi-classical result \eq{drhosc1}, and the dashed line (blue) the
approximation \eq{delrhorad} (for $D$=1) valid for small $x$ values.
{\it Lower panel:} Tests of relations \eq{lvt} and \eq{tautau1}
between the quantum-mechanical densities for the same system. Solid
line (red): $\delta\tau(x)$, dashed line (blue): $-\delta\tau_1(x)$,
dotted line (black): r.h.s.\ of \eq{lvt}.
[Corrected figure from \cite{rb}.]
}
\end{figure}

A numerical example was given in \cite{rb} for the
quartic oscillator in one dimension
\be
V(x)=x^4\!/4\,.
\label{Vx4}
\ee
Unfortunately, an error occurred in the drawing of Fig.\ 1
in \cite{rb}; the present \fig{1dx4drho} is its corrected version.
In the upper panel, it is seen that the semiclassical
approximation \eq{drhosc1} for $\delta\rho(x)$ agrees
very well with the quantum result, and in the lower panel
the relations \eq{lvt} and \eq{tautau1} between the quantum
results are seen to be well fulfilled. The only sizable
deviations occur very near the classical turning point, as
expected.

We emphasize that the Friedel oscillations near the surface
are dominated by the primitive ``+'' orbit (with $k=0$). Its
contribution diverges, however, since its running time
$T_+^{(0)}(\lambdab,x)$ tends to zero there. This divergence
can be remedied in the WKB-type linear approximation to the
potential which we discuss in \cite{nlvt} for smooth
potentials, or by the short-time propagator for hard-wall
potentials (i.e., billiard systems) discussed in \sec{secfriedbil}.
First we will, however, examine the strictly linear potential
for which the WKB approximation is exact.

\subsubsection{The linear potential}
\label{secscllin}

In \cite{nlvt}, we give the exact quantum-mechanical
densities for the one-dimensional potential $V(x)=ax$. Although
this potential does not bind any particles, its density close
to the turning point will be of use in \sec{secfriedlin}.
Here we give its semiclassical analysis.

Since a particle cannot be bound in this
potential, the only closed classical orbit starting from a point
$x$ is the primitive orbit ``+'' ($k=0$) going to the turning
point $x_\lambda=\lambdab/a$ and back to $x$. Its action is
\bea
&&\hspace{-.5cm}
S_+(x)=S_+^{(0)}(x) = 2\!\int_x^{x_\lambda}\!\! p(\lambdab,x)\,\d x
                    = \frac{4\sqrt{2m}}{3a}\,\,(\lambdab-ax)^{3/2}\nonumber \\
  &&\hspace{.5cm}  = \hbar\,\frac43\,|z_\lambda|^{3/2}=\hbar\,2\zeta_\lambda,
\label{splinap}
\eea
where the last equalities make use of the quantities defined as \be
\sigma = \left(\frac{2m}{\hbar^2a^2}\right)^{\!1/3},
\label{sigma}
\ee
and
\be
z_\lambda = \sigma(ax-\lambda)\,,\qquad
\rho_0 = 2\left(\frac{2ma}{\hbar^2}\right)^{\!1/3} = 2\,\sigma a\,.
\label{zmu}
\ee
Using \eq{drhosc} for $D=1$, we
obtain the semiclassical contribution of this orbit to the spatial
density [cf.\ Eq. (23) of \cite{rb} with $\sigma=+$, $k=1$]
\be
\delta\rho(x) = -\frac{a}{2\pi}\,\frac{1}{(\lambdab-ax)}\,
                \cos\left[\frac{1}{\hbar}\,S_+(x)\right],
\label{drholinsc}
\ee
which is identical to the asymptotic expression for
the exact quantum-mechanical result \cite{nlvt}. Thus, the orbit ``+'' creates
the Friedel oscillations. Using the LVT \eq{lvt}
and $Q=-1$ in \eq{dtau1sc}, we obtain immediately the expression
for the kinetic-energy densities
\be
\delta\tau(x) = -\delta\tau_1(x) = -\frac{a}{2\pi}\,
                \cos\left[\frac{1}{\hbar}\,S_+(x)\right],
\label{dtaulinsc}
\ee
which is identical to the asymptotic quantum result \cite{nlvt}.
The expression \eq{drholinsc} diverges at
the classical turning point $x_\lambda$. To avoid this divergence
one has to use the exact expressions \cite{nlvt},
which can be considered as the regularized contributions of the
primitive ``+'' orbit near the turning points.

\subsubsection{The 1-dimensional box}

For the one-dimensional box of length $L$, Eq.\ \eq{drhosc1} has
to be modified by omitting the phase factor $(-1)^k$, since each
turning point gives two units to the Morse index. Using
$\rho_{\text{TF}} =2/\pi \sqrt{2m\lambda_{\text{TF}}/\hbar^2}$
and summing over all $k$, one finds that it reproduces exactly
the quantum-mechanical $\rho(x)$ in the large-$N$ limit,
so that the semiclassical approximation here is
asymptotically exact.

\subsection{$D>1$ dimensional potentials with spherical symmetry}
\label{secrad}

In this section we discuss potentials in $D>1$ with spherical symmetry, so
that $V(\bfr)=V(r)$ depends only on the radial variable $r=|\bfr|$. The
particle number $N$ is chosen such that energy levels with angular-momentum
degeneracy are filled so that all spatial densities, too, depend only on $r$.
In such systems, the two kinds of oscillations discussed in \sec{secsep}
can always be separated clearly in the central region $r\simeq 0$.
Indeed, this behavior is explained by the fact that the angular momentum
of the orbits is conserved. Therefore, the shape of a closed orbit whose
starting point $r$ approaches the center of the potential tends to become
flattened and concentrated near a radial periodic orbit. Thus, close
to the center there are only two types of non-periodic orbits: Firstly,
the {\it radial} orbits of the same types ``$+$'' and ``$-$'' as discussed
for the one-dimensional case, with opposite momenta $\bfp=-\bfp'$, leading
to the same kind of oscillations that we know for $D=1$. Secondly,
{\it non-radial} orbits which near $r=0$ have almost equal momenta
$\bfp\simeq\bfp'$, so that they become nearly periodic.

Semiclassically, the two types of radial and non-radial NPOs are
responsible precisely for the two kinds of oscillations which we
described in \sec{secsep}. The regular short-ranged oscillations,
denoted $\delta_{\text{r}}\rho(r)$ etc., can be attributed to the
radial ``+'' and ``$-$'' orbits. The long-ranged irregular
oscillations, denoted $\delta_{\text{irr}}\rho(r)$ etc., must be
attributed to the non-radial NPOs: these lead to slow oscillations
because their actions are almost independent of the starting point
near $r=0$.

The contributions of the radial NPOs in radially
symmetric systems has already been anticipated in \cite{rb};
they will be discussed in the following section. In particular,
like for $D=1$, the primitive ``+'' orbit is seen to be solely
responsible for the Friedel oscillations near the surface of
a $D>1$ dimensional spherical system.

Non-Radial orbits can only occur if there exist classical trajectories
which intersect themselves in a given point $\bfr$. As is
well known from classical mechanics, such orbits do not exist
in isotropic harmonic oscillators (and in the Coulomb potential).
This explains the fact that no irregular long-ranged oscillations
are found in the densities of harmonic oscillators \cite{bm}
(or, trivially, in any one-dimensional potential).

We emphasize that for $D=2$, all closed NPOs are isolated except
if they start at $r=0$, in which case they form degenerate families
due to the radial symmetry (cf.\ \sec{secregul}). In $D>2$ dimensions,
however, also the non-radial NPOs starting at $r>0$ have
continuous rotational degeneracies. For the corresponding families
of orbits, the Van Vleck determinant ${\cal D}_\gamma$ in the
semiclassical Green function \eq{sclgreen} becomes singular at all
points $r$. This divergence can be removed \cite{strut} by going one
step back in the derivation of \eq{sclgreen}. In the convolution
integral for the time-dependent propagator, one has to perform a
sufficient number of intermediate integrals exactly rather than in
the stationary-phase approximation (for details, see \cite{strut}
where this was done to obtain the trace formula \eq{trf} for systems
with continuous symmetries). As a result, the semiclassical amplitudes
of the degenerate families of orbits are of lower order in $\hbar$
than for isolated orbits and thus have a larger weight. In our present
case, the $\hbar$ dependence of the ratio of amplitudes between the
irregular and the regular oscillations e.g.\ in the particle
density becomes:
\begin{equation}
\frac{|\delta_{\text{irr}}\rho(r)|}{|\delta_{\text{r}}\rho(r)|}
      \propto \hbar^{-(D-2)/2}. \qquad (D>1)
\label{family}
\end{equation}
The same ratio holds also for the other spatial densities.
This can be seen, e.g., in \fig{osc} for the spherical billiard
in $D=3$, where the amplitude of the irregular oscillations is
larger than that of the radial oscillations (except near $r=0$).
In passing, we note that for spherical billiards with radius $R$,
the energy dependence of the semiclassical results scales with
the dimensionless variable $p_\lambda R/\hbar$, and the ratio
\eq{family} becomes $|\delta_{\text{irr}}\rho/\delta_{\text{r}}\rho|
\propto [p_\lambda R/\hbar]^{(D-2)/2}$.

It should be stressed that the separation of two classes of NPOs and
hence the two types of oscillations is not possible in systems in $D>1$
dimensions without radial symmetry. This will be illustrated in \sec{secnonrad}.

A further complication in systems with $D>1$ is that the NPOs can
undergo bifurcations under the variation of the starting point $r$. At
these bifurcations, new NPOs or POs are created. This is discussed
extensively in a publication \cite{circ} on the two-dimensional
circular billiard. For this system, a complete classification of all
NPOs could be made and analytical expressions for their actions and Van
Vleck determinants have been derived.

\subsubsection{Contributions of the radial orbits: Earlier results}
\label{secradlin}

Recall that since all radial NPOs fulfill $\bfp'=-\bfp$, they have
$Q(\lambdab,r)=-1$ under the sum in \eq{dtau1sc}. Therefore we
immediately obtain the semiclassical relation \cite{rb}
\begin{equation}
\delta_{\text{r}}\tau_1(r) \simeq -\delta_{\text{r}}\tau(r)\,.
\label{tautau1r}
\end{equation}
Indeed, this was found to be fulfilled, sufficiently far from the
turning point, for all quantum systems discussed in \sec{secexa}.

In order to derive some of the other forms of local virial
theorems discussed
in \sec{secexa}, it is important to notice the action of the
differential operator $\nabla$ on the semiclassical density in
\eq{drhosc}. The contributions of leading-order in $\hbar$ (i.e.,
the terms of the largest {\it negative} power of $\hbar$) come
from the phase $\Phi(\lambdab,{\bf r})$ given in \eq{phase}.
From the canonical relations \eq{pcanon} we find
\be
\nabla\,e^{i\Phi(\lambdab,\bfr)} = \frac{i}{\hbar}\,(\bfp'-\bfp)\,
                                   e^{i\Phi(\lambdab,\bfr)},
\label{nablaphas}
\ee
and
\be
\nabla^2\,e^{i\Phi(\lambdab,\bfr)} = -\frac{1}{\hbar^2}\,(\bfp'-\bfp)^2\,
                                     e^{i\Phi(\lambdab,\bfr)},
\label{lapphas}
\ee
which occurs for each NPO under the summation in \eq{drhosc}.
For the radial orbits, one therefore obtains with \eq{pclass}
the following differential equation for $\delta_{\text{r}}\rho(r)$, which
was already given in \cite{bm}:
\be
- \frac{\hbar^2}{8m}\,\nabla^2\delta_{\text{r}}\rho(r) \simeq
  [\lambdab-V(r)]\,\delta_{\text{r}}\rho(r)\,.
\label{lapeqv}
\ee
For small distances $r$ from the center so that $V(r)\ll\lambdab$,
\eq{lapeqv} becomes the universal Laplace equation
\be
- \frac{\hbar^2}{8m}\,\nabla^2\delta_{\text{r}}\rho(r)
\simeq \lambdab\,\delta_{\text{r}}\rho(r)\,,
\label{lapeq}
\ee
which was obtained asymptotically from the quantum-mechanical
densities of isotropic harmonic oscillators in \cite{bm}. It has the general solution
\begin{equation}
\delta_{\text{r}}\rho(r) = (-1)^{^{M_{\text{s}}\!-1}}\frac{m}{\hbar\,T_{\text{r1}}(\lambdab)}
                           \left(\frac{p_\lambda}{4\pi\hbar r}\right)^{\!\nu}
                           \!\!J_\nu(2rp_\lambda/\hbar)\,.
\label{delrhorad}
\end{equation}
Here $J_\nu(z)$ is a Bessel function with index $\nu=D/2-1$, $M_{\text{s}}=M+1$ is the
number of filled main shells \cite{noteM}, $T_{\text{r1}}$ is the period of
the primitive radial full oscillation and $p_\lambda=(2m\lambdab)^{1/2}$
is the Fermi momentum. The normalization of \eq{delrhorad} cannot be
obtained from the linear equation \eq{lapeq}; we have determined
it from the calculation presented in \sec{secregul}. For harmonic
oscillators, where $T_{\text{r1}}=2\pi/\omega$, equation
\eq{delrhorad} becomes identical with the result in \cite{bm},
Eq.\ (69), that was derived from quantum mechanics in the large-$N$ limit.

The quantity $\delta_{\text{r}}\rho(r)$ can also be calculated directly
from \eq{drhosc}, including only the radial NPOs. The summation over
their repetitions goes exactly like in the one-dimensional case done
in \cite{rb}, except for the evaluation of the determinant
${\cal D}_\perp$. This determinant becomes singular at $r=0$ due to
the continuous degeneracy of the ``+'' and ``$-$'' orbits: the point
$r=0$ is a caustic point for all radially symmetric systems with
$D>1$. The regularization of this singularity, leading precisely to
the result $\eq{delrhorad}$, is discussed in \sec{secregul} below.

\subsubsection{Isotropic harmonic oscillators in $D$ dimensions}
\label{sechoscl}

We now investigate the densities in the isotropic harmonic oscillator
(IHO) potential in $D$ dimensions defined as
\be
V(r)=\frac{m}{2}\,\omega^2r^2,\qquad r=|\bfr|\,,\qquad \bfr\in\mathbb{R}^D.
\label{vho}
\ee
First we mention the well-know fact that in IHO potentials
with arbitrary $D>1$, all orbits with nonzero angular momentum
are periodic, forming ellipses which may degenerate to circles or radial
librations. Hence the only NPOs are the radial orbits ``+'' and ``$-$''.
Since we just have seen that in the leading-order semiclassical approximation,
$\delta_{\text{r}}\tau_1(r)=-\delta_{\text{r}}\tau(r)$, it follows that
$\delta\xi(r)=0$ to leading order like for $D=1$, thus explaining the
smooth behavior of $\xi(r)$ for IHOs \cite{bm}.

For the IHO potentials, the transverse determinant ${\cal D}_{\bot}$ can
be easily computed. It is diagonal and reads
\begin{equation}
|{\cal D}_{\bot}(\lambdab,r)|=\left[\frac{m \lambdab}{r p(\lambdab,r)}\right]^{D-1},
\end{equation}
which does not depend on the type and the repetition number $k$ of the orbit.
Following \eq{drhosc} and \cite{rb}, we compute $\delta\rho(r)$
as a sum over the contributions of the ``$+$'' and ``$-$'' orbits,
which is given by
\bea
\delta\rho(r) & = & \frac{4m \hbar}{(2\pi\hbar )^{\frac{D+1}{2}}}
                    \frac{1}{p(\lambdab,r)}\left[\frac{m\lambdab}{r p(\lambdab,r)}\right]
                    ^{\!\frac{D-1}{2}}\nonumber\\
    & & \!\!\!\times \!\!\sum_{k=0,\pm}^\infty\!\! \frac{\cos\!\left[S_{\pm}^{(k)}(\lambdab,r)-(D+1)
        \frac{\pi}{4}\!-\mu_{\pm}^{(k)}\frac{\pi}{2}\right]}{T_{\pm}^{(k)}(\lambdab,r)}.~~~~~~~
\label{rho_HO}
\eea
Here we have used the analytical form of the actions and periods
\bea
S_{\pm}^{(k)}(\lambdab,r) & = & (2k+1) \frac{\pi\lambdab}{\omega} \mp r p(\lambdab,r)\nonumber\\
                            & & \mp\, \frac{2\lambdab}{\omega}\,
                                \arcsin\left(\frac{m\omega r}{p_\lambda}\right)\,,\\
T_{\pm}^{(k)}(\lambdab,r) & = & (2k+1) \frac{\pi}{\omega}\, \mp \frac{2}{\omega}\,
                                \arcsin\left(\frac{m\omega r}{p_\lambda}\right).~~~
\eea
We compute the Morse indices following Gutzwiller \cite{gutz1}.
Each turning point contributes a phase of $\pi/2$. Besides
we evaluate the number of extra conjugate points including their
multiplicities depending on the dimension, contributing a
phase $\pi(D-1)/2$ each (they are most easily determined from
the propagator of the harmonic oscillator in the time
representation). The final result for the Morse indices is
\begin{equation}
\mu_{+}^{(k)}=2 k D+1\,, \qquad \mu_{-}^{(k)}=2 k D +D\,.
\label{Mascorr}
\end{equation}
We note that the equation \eq{rho_HO} is consistent
with results derived in \cite{mue} from the quantum mechanical
density $\rho(r)$.
\begin{figure}
\includegraphics[width=1\columnwidth,clip=true]{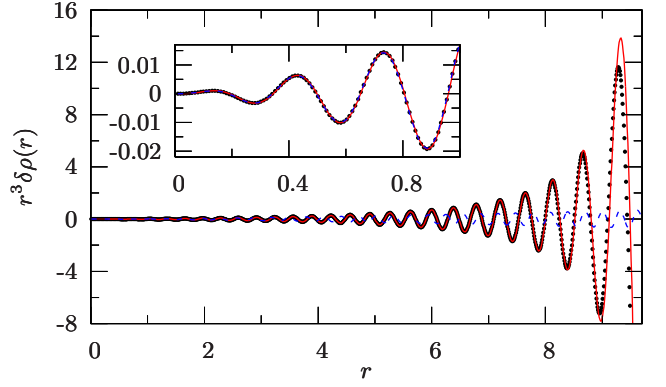}
\caption{\label{fig_2D_HO}
(Color online) Oscillating part of the spatial particle density times $r^3$
for 4$D$ IHO for $N=632502$, i.e.\ with $M=50$ filled shells (units:
$\hbar=m=\omega=1$). Dots are the quantum results. The
solid (red) line is the analytical expression (\ref{rho_HO}) using the
Morse indices given in \eq{Mascorr}, and the dashed (blue) line is the
asymptotic formula (\ref{delrhorad}) valid close to $r=0$.
}
\end{figure}
Fig.\ \ref{fig_2D_HO} shows a comparison of the semiclassical results
\eq{rho_HO} with the exact quantum result for the case $D=4$. We
have multiplied both by a factor $r^3$ since the semiclassical
determinant ${\cal D}_{\bot}$ diverges at $r=0$ which is a caustic
point due to the spherical symmetry. This divergence will be
regularized in the following section.

Using the Morse indices \eq{Mascorr} and for $\lambdab$ the expression
$\lambdab~=~\hbar\omega~[M+(D+1)/2]$ \cite{bm}, we can perform the summation over $k$ in \eq{rho_HO}
analytically for small $r$, like it was done in \cite{rb} for the 1$D$
case. The result then is exactly that given in \eq{delrhorad} with
$T_{\text{r1}}(\lambdab)=2\pi/\omega$, but replacing the Bessel function
$J_\nu(z)$ by its asymptotic expression for large argument $z$, i.e.,
using
\be
J_\nu(z) \quad\rightarrow\quad \sqrt{\frac{2}{\pi z}}
         \cos(z-\nu\pi/2-\pi/4)\,.
\label{Besasy}
\ee

\subsubsection{Regularization close to the center}
\label{secregul}

In this section we compute the contribution of radial NPOs to
the semiclassical particle density close to the center of an arbitrary
potential with radial symmetry. As stressed in the last section,
the semiclassical Green function for $D>1$ is not defined at $r=0$
where ${\cal D}_{\bot}$ diverges. The reason is the caustic that
occurs there: fixing the position of the point $r=r'=0$ does not
uniquely determine a closed orbit (periodic or non-periodic) which
belongs to a continuously degenerate family due to the spherical
symmetry. A standard method to solve this problem is to introduce the
mixed phase-space representation of the Green function close to the
diverging point, as proposed initially by Maslov and Fedoriuk \cite{mf}.

Here we follow more specifically the procedure outlined in
\cite{ruj}. The mixed representation of the Green function can be
approximated in a form analogous to that in the coordinate representation.
This is due to the smoothness of the phase-space torus which implies that
no diverging points can occur simultaneously in position and momentum
(cf.\ \cite{mf}). Following Gutzwiller, we use for every classical
trajectory $\gamma$ an ``intrinsic'' (or local) coordinate system
$\bfr=(r_\parallel,\bfr_\perp)$, where the coordinate $r_\parallel$ is
taken along the trajectory and $\bfr_\perp$ is the vector of all other
coordinates transverse to it; $\bfp=(p_\parallel,\bfp_\perp)$ is the
corresponding system for the momentum.
We next re-write the coordinate-representation of the Green function
as the inverse Fourier transform of the mixed Green function with
respect to the final transverse momentum $\bfp'_\perp$:
\begin{eqnarray}
&&\hspace{-1cm}G_{\text{scl}}(E,\bfr,r'_\parallel,\bfr'_\perp) = \frac{1}{(-2 i \pi \hbar)^{(D-1)/2}}
                                           \sum_\gamma \int \d \bfp'_\perp \nonumber\\
                                        && \hspace{1.cm}  \times~ \Gft_\gamma(E,\bfr,r'_\parallel,\bfp'_\perp)\,
                                           \exp\left(\frac{i}{\hbar}\,
                                           \bfr'_\perp\cdot\bfp'_\perp\right)\!,~~
\label{def_green_ft}
\end{eqnarray}
where the sum is over all classical trajectories $\gamma$ starting at $\bfr$
and ending at $(r'_\parallel,\bfp'_\perp)$ in phase space. Hereby the contribution of the
orbit $\gamma$ to the semiclassical mixed representation of the Green function
is given by \cite{mf}:
\begin{eqnarray}
&&\hspace{-1cm}\Gft_\gamma(E,\bfr,r'_\parallel,\bfp'_\perp)=\alpha_D\, \mathrm{\widehat{\cal D}}_\gamma
                                               (E,\bfr,r'_\parallel,\bfp'_\perp)\nonumber\\
                                            && \hspace{1.cm}\times \exp {\bigg (\frac{i}{\hbar}
                                               \widehat{S}_\gamma
                                               (E,\bfr,r'_\parallel,\bfp'_\perp)-\frac{i\pi}{2}
                                               \widehat{\mu}_\gamma\bigg)},~
\label{def_mix_green}
\end{eqnarray}
where $\widehat{S}$ is the Legendre transform of the action $S$ between
the variables $\bfr'_\perp$ and $\bfp'_\perp$:
\be
\widehat{S}_\gamma(E,\bfr,r'_\parallel,\bfp'_\perp)
       = S_\gamma(E,\bfr,r'_\parallel,\bfr'_\perp) -{\bf r'}_{\!\bot}\cdot{\bf p'}_{\!\bot}\,.
\label{legtra}
\ee
Since in the mixed-representation Green function, we have to evaluate the action
$\widehat{S}$ for radial orbits with {\it fixed} momentum close to the center,
the rotational symmetry in position is removed and $\widehat{G}$ is regular.
The Van Vleck determinant in this representation is
\begin{equation}\label{detp}
\mathrm{\widehat{\cal D}}_\gamma = \frac{m}{|p_\parallel p'_\parallel|^{1/2}}\,
                                    |\mathrm{\widehat{\cal D}}_{\!\bot\gamma} |^{1/2}\,,
\quad \mathrm{\widehat{\cal D}}_{\!\bot \gamma}=  \det \bigg(\dfrac{\partial\bfp_\perp}{\partial\bf
p'_\perp}\bigg) \,,
\ee
and the Morse index becomes
\begin{eqnarray}
\widehat{\mu}_\gamma = \left\{ \begin{array}{ll}
                       \mu_\gamma&\hbox{for positive eigenvalue of}
                       ~\det\bigg(\dfrac{\partial{\bf r'}_{\!\bot}}{\partial{\bf p'}_{\!\bot}}\bigg
)\,,\nonumber\\
\nonumber\\
                       \mu_\gamma+1&\hbox{for negative eigenvalue of}
                       ~\det\bigg(\dfrac{\partial{\bf r'}_{\!\bot}}{\partial{\bf p'}_{\!\bot}}\bigg
)\,.
                               \end{array}\right.\nonumber
\end{eqnarray}
Far from singular points, the evaluation of \eq{def_green_ft} using the
stationary phase approximation (SPA) yields \cite{spa} the standard
semiclassical Green function \eq{sclgreen}.

After performing the $\hbar$ expansion and the
integration over the energy similarly as in \cite{rb}, the oscillating
part of the particle density is given by:
\begin{eqnarray}
\hspace{-1.5cm}
\delta \rho (\bf r) &=& 2 \sum_{\gamma} \text{Im} \bigg \{ \frac{i \alpha_D \hbar }{\pi T}
\int \d{\bf p'}_{\!\bot}
\widehat{G}(\lambdab,{\bf r},r_{\parallel},{\bf p'}_{\!\bot})\nonumber \\
&&\hspace{2.3cm}\times \exp \bigg ( \frac{i}{\hbar} {\bf r}_{\!\bot}
\cdot{\bf p'}_{\!\bot}\bigg) \bigg \}.
\label{drhomix}
\end{eqnarray}
Close to the center of the potential, now, we replace the non-radial
NPOs by the radial ones with $\gamma=$''$\pm$'' orbits with $k$-th repetitions.
For non-periodic orbits in the radial direction $r$ we have
$r_\parallel=r$.
We neglect the higher orders in ${\bf r}_{\!\bot}$,
leading to the following approximations:
\begin{eqnarray}
   &&  |\det (\partial{\bf p}_{\!\bot}/\partial{\bf p'}_{\!\bot})|\;\approx\; 1\,,\nonumber\\
   &&  |p_{_{\parallel}}|\approx |p'_{_{\parallel}}|\;\approx\; p_\parallel(\lambdab,{\bf p'}_{\!\bot})
      := \sqrt{2 m \lambdab-{\bf p'}_{\!\bot}^2}~,\hspace*{0.8cm}  \nonumber\\
   &&  (r,{\bf r}_{\!\bot})\;\approx\; (r,0)\,,\nonumber\\
   &&  \widehat{S}_{\pm}^{(k)}\;\approx\; (k+1/2)\,S_{\text{r1}} \mp 2rp'_{_{\parallel}}\,,\nonumber\\
   &&  T_{\pm}^{(k)}\;\approx\;(k+1/2)\, T_{\text{r1}}\,.
\label{tpmapp}
\end{eqnarray}
Furthermore, we approximate the action $S_{\text{r1}}$ of the primitive
periodic diameter orbit by $S_{\text{r1}}\approx 2 \pi \hbar[M+(D+1)/2]$.
This is exact for IHOs where $S_{\text{r1}}=2\pi\lambdab/\omega$
and $\lambdab~=~\hbar\omega~[M+(D+1)/2]$ can be used \cite{bm};
for arbitrary radial potentials
it corresponds to a radial WKB quantization, whereby $M$ is a
``main shell'' quantum number that has to be suitably chosen
\cite{noteM}. Also, we assume that each eigenvalue of
$\det(\partial{\bf r'}_{\!\bot}/\partial{\bf p'}_{\!\bot})$ is
negative (positive) for the orbits ``$+$'' (``$-$''), leading to
$\widehat{\mu}_{+}^{(k)}=\widehat{\mu}_{-}^{(k)}$. This is again
exact for IHOs; for other radial potentials we have verified its
validity numerically. With these approximations, the sum over the
repetitions of all radial orbits can be performed exactly like in
the previous section. The oscillating part of the
particle density then simplifies to:
\begin{equation}
\delta_{\text{r}} \rho(r) = \frac{(-1)^{M}m}{
                   (2 \pi \hbar)^{D-1} T_{\text{r1}}} \int \d {\bf p'}_{\!\bot}
                   \frac{\cos[2rp_\parallel(\lambdab,{\bf p'}_{\!\bot})/\hbar]}
                   {p_\parallel(\lambdab,{\bf p'}_{\!\bot})}\,.
\end{equation}
The integration has to be taken over half the solid angle in the
$(D-1)$-dimensional transverse momentum space, avoiding a
double-counting of the two orbits. So it is natural
to make a change of variables to dimensionless hyper-spherical
coordinates. Using the integral representation of the Bessel
functions \cite{abro}
\be
\displaystyle J_\nu(z) = \frac{2(z/2)^{-\nu}}{\sqrt{\pi}
                         \Gamma(\nu+1/2)} \int_0^1(1-t^2)^{\nu-1/2}\cos(zt)\, \d t\,,
\ee
we obtain exactly the same result as in \eq{delrhorad}, confirming
its normalization.

We stress that this regularization is only valid near the center, i.e., for
$r\simeq 0$, as can be seen in the example of \fig{fig_2D_HO}, where the result
\eq{delrhorad} is displayed by the dashed line. The reason is that for larger
values of $r$, the approximations \eq{tpmapp} are no longer valid. If one
restricts oneself to the leading contributions of the primitive orbits
``+'' and ``$-$'' with $k=0$, a ``global uniform'' approximation can be
made which interpolates smoothly between the regularized result \eq{delrhorad}
near $r=0$ and the correct semiclassical contributions obtained from
\eq{drhosc} at larger $r$. This uniform approximation is derived and
used in \cite{circ} for the $2D$ circular billiard system which we
briefly discuss in the following section.

\subsubsection{The two-dimensional circular billiard}
\label{secdisc}

The two-dimensional circular billiard, which can be taken as
a realistic model for quantum dots with a large number $N$
of particles, has
been investigated semiclassically in \cite{circ}, where all
its periodic and nonperiodic closed orbits have been classified
analytically. We discuss there also the various bifurcations
at specific values of the radial variable $r$, at which POs
bifurcate from NPOs or pairs of NPOs are born. At these
bifurcations, the semiclassical amplitudes in \eq{drhosc}
-- \eq{dtau1sc} must be regularized by suitable uniform
approximations. We refer to \cite{circ} for the details
and reproduce here some numerical results to illustrate the
quality of the semiclassical approximation.
\begin{figure}[h]
\includegraphics[width=1.15\columnwidth,clip=true]{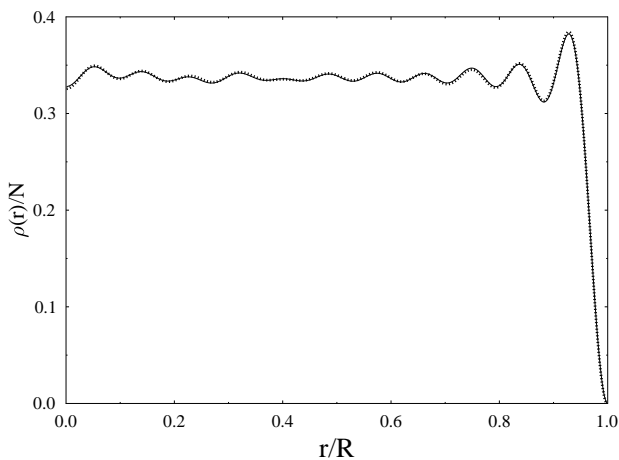}\vspace*{-0.5cm}
\caption{\label{disk606}
Particle density in the two-dimensional disk billiard with radius $R$,
containing $N=606$ particles (units: $\hbar^2\!/2m=R=1$), divided 
by $N$. The solid line is
the quantum result, the dotted line the semiclassical result
with all regularizations (see \cite{circ} for details).
}
\end{figure}

\fig{disk606} shows the total particle density $\rho(r)$ for
$N=606$ particles in the circular billiard. The solid line
gives the quantum result, obtained from \eq{rho} using the
solutions of the Schr\"odinger equation with Dirichlet boundary
conditions, which are given in terms of cylindrical Bessel
functions. The dotted line gives the semiclassical result, obtained
by summing over the $\sim$ 30 shortest NPOs. (Hereby we used the
regularization of the radial ``+'' and ``$-$'' orbits at $r=0$ by
\eq{delrhorad}, that of the primitive ``+'' orbit near $r=R$ by
\eq{friedreg2d} given in \sec{secfriedel2d} below, and uniform
approximations for the bifurcations of some of the non-radial NPOs
as described in detail in \cite{circ}.) We see that, indeed,
a satisfactory approximation of the quantum density can be
obtained in terms of the shortest classical orbits of this system.
\begin{figure}[h]
\includegraphics[width=1.1\columnwidth,clip=true]{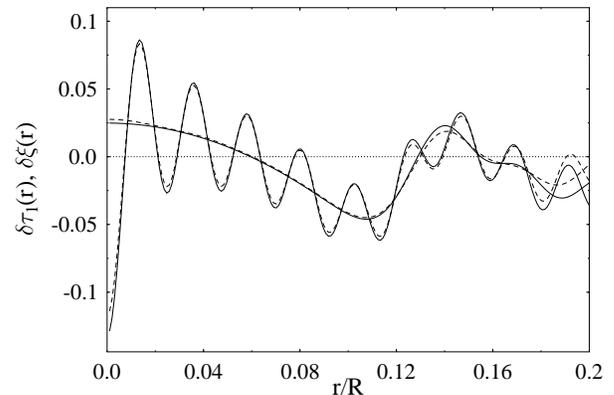}\vspace*{-0.2cm}
\caption{\label{disk9834}
Oscillating parts of kinetic-energy densities, $\delta\tau_1(r)$
(fast oscillations) and $\delta\xi(r)$ (slow oscillations) for
$N=9834$, divided by $N^{5/3}$.
{\it Solid lines:} exact quantum results. {\it Dashed lines:}
semiclassical results and units as in \fig{disk606}.
}
\end{figure}

In \fig{disk9834} we demonstrate explicitly the contributions
of non-radial NPOs to the kinetic-energy densities $\tau_1(r)$ and
$\xi(r)$ close to the center, calculated as in \fig{disk606} but for
$N=9834$ particles. We clearly see that $\delta\xi(r)$ is not smooth;
its slow, irregular oscillations are due to non-radial NPOs which
have the form of polygons with $2k$ reflections ($k=1,2,\dots$) at the
boundary and one corner at a point $r$ close to the center. The first
$k_{\text{max}}=20$ of them were included with the appropriate regularization
at $r=0$ where they are degenerate with the $k$-th repetitions of
the diagonal PO (see \cite{circ} for details). The agreement between
quantum and semiclassical results is again satisfactory; the discrepancy
that sets on for $r\simg 0.18$ is due to the missing of more complicated
non-radial orbits. The quantity $\delta\tau_1(r)$, on the other hand,
clearly exhibits both kinds of oscillations according to \eq{dtau1dec}:
the slow irregular part, which is identical with $\delta\xi(r)$,
is modulated by the regular fast oscillations due to the radial orbits.

\subsection{$D>1$ dimensional systems without continuous symmetries}
\label{secnonrad}

In $D>1$ dimensional systems without continuous symmetries, it is
in general not possible to find the classical orbits analytically.
As in POT, the search of closed orbits must then be done numerically.
A practical problem in such systems is also that the densities
as functions of $D$ coordinates are not easily displayed.
For tests and comparisons of various approximations or of the
local virial theorems, we have to resort to taking suitable
one-dimensional cuts (i.e., projections) of the densities.
In the following paragraph, we discuss a class of integrable
billiard systems, in which all closed classical orbits can
easily be found and their semiclassical contributions to the
densities can be analytically obtained. These are
$D$-dimensional polygonal billiards that tessellate the full
space under repeated reflections at all borders. We illustrate
the method for the example of a rectangular billiard.
Although this does not correspond to any physical system
(unless experimentally manufactured as a rectangular quantum
dot with many electrons), it is a useful model without
spherical symmetry that allows for analytical calculation 
of the classical orbits and their properties.

\subsubsection{Billiards tessellating flat space:
               the rectangular billiard}
\label{secrect}

For billiards, classical trajectories are straight lines which are
reflected at the boundary according to the specular law. Let us 
consider a two-dimensional billiard that tessellates the plane, such
as the rectangular billiard shown in \fig{fig_tab}. Choose 
a trajectory starting at a point $P$, reflected at the point $R_0$ and
reaching the point $P_1$. Now reflect the boundary at a side containing
the point $R_0$, the image $R_0P_1'$ of the segment $R_0P_1$ gives the
straight line $PP_1'$. The next portion of the trajectory after
reflexion in $R_1$, can be found by reflecting the new billiard at the
side containing $R_1'$. This process can be repeated until the
trajectory ends. To get the closed trajectories at $P$ we have to
compute all images of $P$ in the images of the billiard. Now a
straight line joining $P$ and an image of $P$ gives a closed orbit.
Thus, constructing all images of $P$ by simple geometry, enables one to
compute all trajectories and their related initial and final momenta
for $D$-dimensional polygonal billiard that fills the $D$-dimensional
Euclidian space. 
Note that the Jacobian $\displaystyle {\cal D}_{\bot}$ for these systems is 
easily computed and equals $(p/L_{\text{NPO}})^{D-1}$ where $L_{\text{NPO}}$
is the length of the orbit.
\begin{figure}[h]\vspace{-0.3cm}
\includegraphics[width=0.95\columnwidth,clip=true]{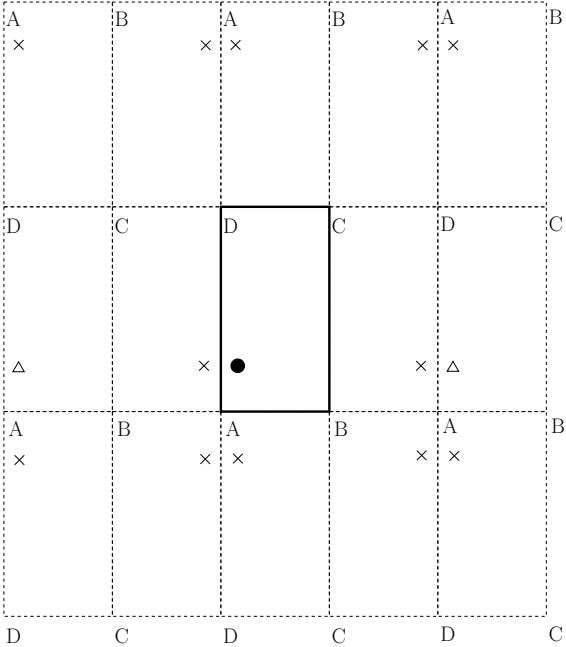}\vspace{-0.3cm}
\caption{\label{fig_tab}
Images (triangles and crosses) of a point $P(x,y)$ (full circle)
for a rectangular billiard. Joining by a straight line the full
circle to a cross gives a non-periodic orbit whereas joining to a
triangle gives a periodic orbit.
}\vspace{-0.2cm}
\end{figure}

We illustrate this method for the case of a 2$D$ rectangular billiard
with side lengths $Q_x$ and $Q_y$. There are four types of images of
$P(x,y)$; one leading to POs and three (labeled by the index a, b and c)
leading to NPOs (see Fig.\ref{fig_tab}). Table \ref{table1} lists the
basic ingredients to compute the spatial densities, using
$L(x,y)=2 \sqrt{x^2+y^2}$ and
\be
f(x,y,\mu) = \frac{4 \hbar p_\lambda^{1/2}}{[2 \pi\hbar L(x,y)]^{3/2}}
             \cos\left[\frac{p_\lambda L(x,y)}{\hbar}-\frac{3 \pi}{4} -\mu \pi\right]\!,
\ee
with $p_\lambda=(2m\lambdab)^{1/2}$. From \eq{drhosc}, \eq{dtau1sc} for $D=2$ we obtain
\begin{eqnarray}
\delta \rho(x,y) & = & \sum_{k_x,k_y=-\infty}^{\infty}\; \sum_{l=\text{a},\text{b},\text{c}}
                       \delta \rho_{l}(x,y)\,,\label{rho_rect}\\
\delta \tau_1(x,y) & = & \sum_{k_x,k_y=-\infty}^{\infty}\; \sum_{l=\text{a},\text{b},\text{c}}
                         \delta \tau_{1l}(x,y)\,,
\label{tau1_rect}
\end{eqnarray}
where the partial contributions $\delta\rho_l(x,y)$ and
$\delta\tau_{1l}(x,y)$ for the orbits of types $l$
= a, b, and c are given in \tab{table1}. $\delta \tau(x,y)$ is
obtained from (\ref{rho_rect}) using the LVT (\ref{lvt}).
\begin{figure}[h]\vspace*{-.3cm}
\includegraphics[width=1.02\columnwidth,clip=true]{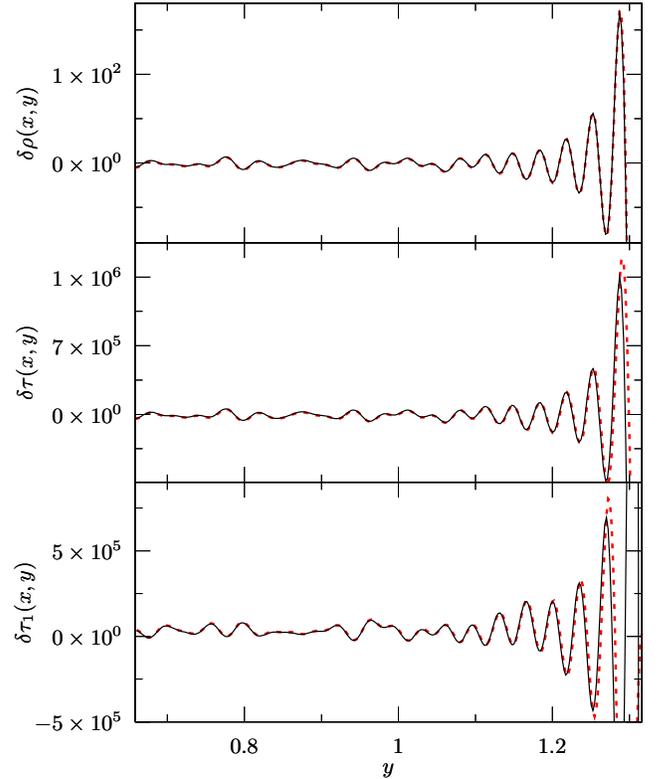}
\vspace*{-0.9cm}
\caption{\label{fig_dens_rect}
(Color online) Oscillating part of the spatial densities for a
rectangular billiard with sides $Q_x=2^{1/4}$ and $Q_y=3^{1/4}$ for
$N=2000$ along the line $x=Q_x/2$ (units: $\hbar^2=2m=1$). Solid
(black) lines are the semiclassical results using \eq{rho_rect},
\eq{tau1_rect} and \eq{lvt}; dashed (red) lines are the
quantum-mechanical results.
}
\end{figure}
\begin{figure}[h]\vspace*{-.3cm}
\includegraphics[width=1\columnwidth,clip=true]{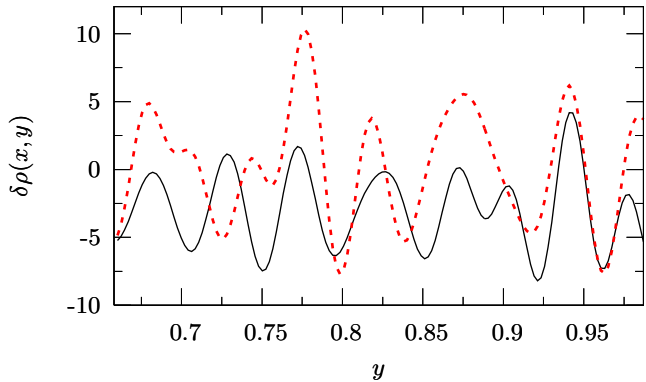}
\vspace*{-0.7cm}
\caption{\label{fig_dens_rect2}
(Color online) Same system as in \fig{fig_dens_rect}. Here, selected
contributions to \eq{rho_rect} of the primitive NPOs with $k=0$ are
shown. The full (black) line gives the contributions of the primitive
self-retracing orbits, and the dashed (red) line that of all other
primitive orbits.
}
\end{figure}

We now present numerical results for the rectangular billiard with side 
lengths $Q_x=2^{1/4}$, $Q_y=3^{1/4}$ (units: $\hbar^2=2m=1$),
containing $N=2000$ particles.
In \fig{fig_dens_rect} we show the quantities $\delta\rho$ (top),
$\delta\tau$ (center) and $\delta\tau_1$ (bottom) as functions of
$y$ with fixed $x=Q_x/2$. Dashed lines are the quantum-mechanical
results, solid lines the semiclassical ones using \eq{rho_rect},
\eq{tau1_rect} and \eq{lvt}. We see that summing over all orbits yields
very good agreement, except close to the boundary where the Friedel
oscillations were not regularized.

In \fig{fig_dens_rect2} we display selected contributions of some of
the primitive orbits $(k=0)$ to the particle density $\delta\rho(x,y)$.
The solid line gives the contribution of self-retracing orbits with
$\bfp=-\bfp'$, and the dashed line that of the other primitive NPOs.
It is evident that no clear separation of regular short-ranged
and irregular long-ranged oscillations can be made here.

\section{Regularization near surface}
\label{secsurf}

As we have pointed out in the previous section,
the semiclassical approximation of density oscillations in
terms of classical orbits breaks down near the classical
turning point due to the diverging amplitude of the
primitive ``+'' orbit (with $k=0$) which close to the
surface is responsible for the Friedel oscillations.
In order to regularize this diverging amplitude, different
techniques must be used for smooth potentials and for
billiards with reflecting walls.

\subsection{Smooth potentials}
\label{secfriedlin}

In smooth potentials $V(\bfr)$, the divergence can be regularized
by linearizing the potential near the classical turning points,
as it is done in the standard WKB approximation \cite{wkb}. In
the surface region close to a turning point, the exact results for
linear potentials given in \cite{nlvt} can then be used.
We demonstrate this first for the one-dimensional case, and then
illustrate it also for potentials in $D=3$ with spherical symmetry.

\subsubsection{Linear approximation to a smooth 1$D$ potential}
\label{seclinap}

We start from an arbitrary smooth binding potential $V(x)$ and
approximate it linearly around the turning point $x_\lambda$
defined by $V(x_\lambda)=\lambdab$. Without loss of generality,
we assume $x_\lambda>0$. Expanding  $V(x)$ around $x_\lambda$ up
to first order in $x-x_\lambda$, we get the approximated potential
\be
{\widetilde V}(x) = \lambdab + a\,(x-x_\lambda)\,,\qquad
                a = V'(x_\lambda)>0\,.
\label{vlinap}
\ee
 We can
therefore apply the results of \cite{nlvt}. The oscillating part of
the density near the turning point then becomes
\bea
&& \hspace{-1.9cm}
\delta\rho_{\text{lin}}(x) = \rho_0\bigg\{[\Ai'(z_\lambda)]^2-z_\lambda\Ai^2(z_\lambda)\nonumber\\
&& \hspace{.5cm}      -\frac{1}{\pi}\sqrt{-z_\lambda}\,\Theta(x_\lambda-x)\bigg\},
\label{drholinap}
\eea
where the last term is the subtracted TF part and
\be
\rho_0 = 2\left(\frac{2ma}{\hbar^2}\right)^{\!1/3}\!\!, \qquad
z_\lambda = \frac{\rho_0}{2}\,(x-x_\lambda)\,.
\label{linap}
\ee
The oscillating parts of the kinetic-energy densities $\tau(x)$ and
$\xi(x)$ become in the same approximation
\bea
&&\hspace{-1.2cm}
\delta\tau_{\text{lin}}(x) = \frac{2a}{3}\bigg\{\Ai(z_\lambda)\Ai'(z_\lambda)
                      -z_\lambda [{\rm Ai}'(z_\lambda)]^2 \nonumber \\
 && \hspace{.8cm}     +z_\lambda^2 {\rm Ai}^2(z_\lambda)
                      -\frac{1}{\pi}|z_\lambda|^{3/2}\Theta(x_\lambda-x)\bigg\},
\label{dtaulinap}\\
&&\hspace{-1.2cm}
\delta\xi_{\text{lin}}(x)   =  -\frac{a}{3}\bigg\{\Ai(z_\lambda)\Ai'(z_\lambda)
                        +2z_\lambda [{\rm Ai}'(z_\lambda)]^2 \nonumber \\
                    &&  \hspace{.8cm}    -2z^2_\lambda {\rm Ai}^2(z_\lambda)
                        +\frac{2}{\pi}|z_\lambda|^{3/2}\Theta(x_\lambda-x)\bigg\}.
\label{dxilinap}
\eea
In the next step, we introduce uniform linearized approximations,
in which the argument $z_\lambda$ in \eq{drholinap}, \eq{dtaulinap},
and \eq{dxilinap} is not as given in \eq{linap}, but replaced by
\be
{\widetilde z}_\lambda = -\left[3S_+(x)/4\hbar\right]^{2/3},
\label{zuni}
\ee
where $S_+(x)$ is the correct action of the ``+'' orbit for the given
potential $V(x)$. On can show that this relation is exact for the
linear potential; it is uniform for other smooth potentials in that
it holds locally at the turning point and yields the correct phase
of the oscillation at all other distances from the turning point.
\begin{figure}[t]
\includegraphics[width=1\columnwidth,clip=true]{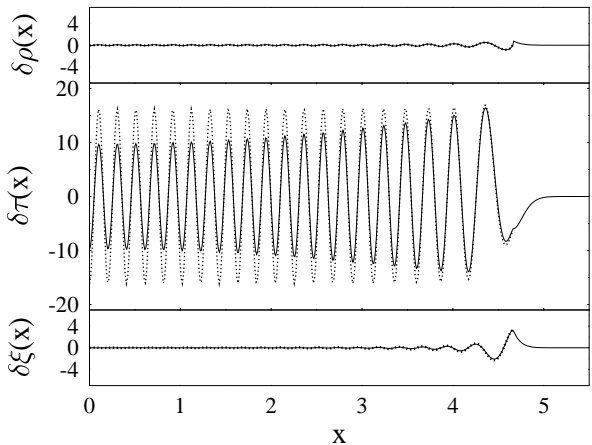}
\vspace*{-0.6cm}
\caption{\label{x4friedun}
Oscillating parts of densities for the quartic potential \eq{Vx4}
with $N=40$ particles (units $\hbar=m=1$), shown on the same scale.
{\it Solid lines:} exact quantum-mechanical results, {\it dotted
lines:} uniform linearized approximations \eq{drholinap}, \eq{dtaulinap}
and \eq{dxilinap} with the argument ${\widetilde z}_\lambda$ given in
\eq{zuni}.
}
\end{figure}

Figure \ref{x4friedun} shows numerical results for these uniform
approximations for the quartic oscillator \eq{Vx4} with $N=40$
particles, compared to the exact quantum results. We see that the
uniform linearized approximation reproduces very well the Friedel
oscillations near the turning point in all three densities.
The phase of the oscillations is seen to be correct at all
distances. The amplitudes are not exact in the asymptotic region,
i.e., near $x=0$. This is not surprising, since the contributions
of all ``$-$'' orbits and those of the ``+'' with $k>0$ are
missing in this approximation. We see that $\delta\xi(x)$ vanishes
inside the system, as expected from the semiclassical leading-order
result on the r.h.s.\ of \eq{tautau1}.
However, near the turning point, where the semiclassical approximation
breaks down, the magnitude of $\delta\xi(x)$ is comparable to --
and for the quartic potential even larger than -- that of $\delta\rho(x)$.
(Note that all three density oscillations are shown on the same vertical
scale.)

\subsubsection{Linear approximation to smooth radially symmetric potentials in $D>1$}
\label{radlinap}

We now start from an arbitrary smooth binding potential with radial symmetry,
$V(\bfr)=V(r)$, $r=|\bfr|$, in $D>1$ dimensions. As above, we replace it by its
linear approximation around the turning point $r_\lambda$ analogously to \eq{vlinap}:
\be
{\widetilde V}(\bfr) = \lambdab + \bfa \cdot (\bfr-\bfr_\lambda)\,,\qquad
                       \bfa = \nabla V(r_\lambda)\,.
\label{vlinaprad}
\ee
Due to the spherical symmetry of $V(r)$, all components of the vector
$\bfa$ have the same magnitudes:
\be
\bfa = a\,\bfr_\lambda/r_\lambda\,,\qquad
a = V'(r_\lambda)\,.
\ee
Therefore, we may choose the radial variable $r$ along any of the
Cartesian axes $x_i$, and the results
\bea
\rho(x_i) & = & -\frac{1}{48\pi}\,\rho_{i0}^3\,
                \bigg \{\Ai(z_{i\lambda})\Ai'(z_{i\lambda})+2z_{i\lambda}
              [{\rm Ai}'(z_{i\lambda})]^2\nonumber\\
          & & \hspace{1.65cm} -2z^2_{i\lambda} {\rm Ai}^2(z_{i\lambda})\bigg\},
              \quad (D=3)~~~
\label{rholin3}
\eea
where $\rho_{i0}=2\sigma_ia_i$,
taken from \cite{nlvt} for the linear potential with $D>1$
apply with the replacements $x_i\rightarrow r$, $z_{i\lambda}\rightarrow
\sigma(ar-\lambda)$. As in \sec{seclinap} for $D=1$, we may then
subtract their ETF contribution. Finally we introduce the uniform
approximation to their oscillating parts near the surface with the
argument \eq{zuni} expressed in terms of the action $S_+(r)$ of the
primitive radial ``+'' orbit of the given radial potential $V(r)$:
\be
{\widetilde z}_\lambda = -\left[3S_+(r)/4\hbar\right]^{2/3}.
\label{zunirad}
\ee

\begin{figure}[h]
\includegraphics[width=1.09\columnwidth,clip=true]{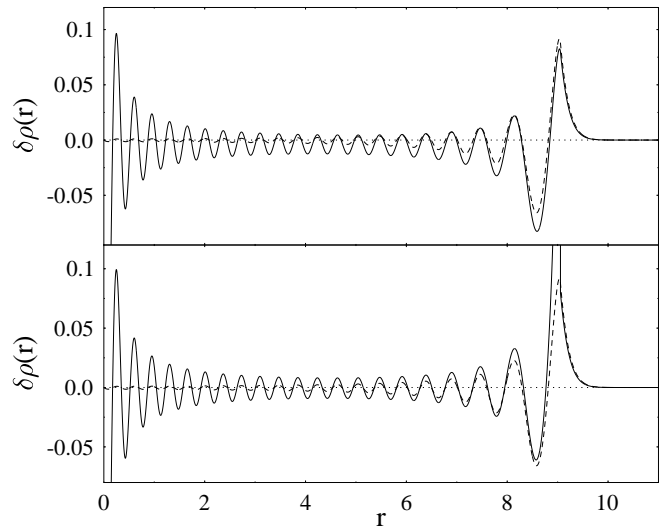}\vspace*{-0.3cm}
\caption{\label{3hoairy}
Oscillating part of particle density for the $3D$ IHO with $N=22960$
particles ($M_{\text{s}}=40$) (units $\hbar=m=\omega=1$).
{\it Solid lines:} exact results, {\it dashed
lines:} uniform linearized approximation (\ref{rholin3})from 
\cite{nlvt} with argument ${\widetilde z}_\lambda$ given in \eq{zunirad}.
{\it Upper panel:} smooth part in $\delta\rho(r)$ taken as
TF density, {\it lower panel:} smooth part in $\delta\rho(r)$
taken as ETF density.
}
\end{figure}
In \fig{3hoairy} we show numerical results for this approximation
for the 3-dimensional IHO with $M_{\text{s}}=40$ occupied shells. The upper
panel shows the exact result for $\delta\rho(r)$ (solid line),
whereby only the TF approximation was used for its smooth part:
$\delta\rho(r)=\rho(r)-\rho_{\text{TF}}(r)$. We notice that the oscillations
in the interior are not symmetric about the zero line, which is due to
smooth errors in the TF density. In the lower panel, the ETF
corrections have been included in $\delta\rho(r)$; now the oscillations
are symmetric about zero. The price paid for this is that $\delta\rho(r)$
diverges at the classical turning point. The uniform linear approximation \eq{rholin3}
with the argument \eq{zunirad}, shown in both panels by
the dashed lines, reproduces well the Friedel oscillation near the
surface. In the interior, it fails due to the missing contributions
of the repetitions ($k>0$) of the ``+'' and of all ``$-$'' orbits.
Once more, these results demonstrate that the Friedel oscillations
near the surface are semiclassically explained by the primitive
``+'' orbit alone. Its diverging amplitudes according to \eq{drhosc}
must, however, by regularized by the uniform linear approximation.
\begin{figure}[h]
\includegraphics[width=1.18\columnwidth,clip=true]{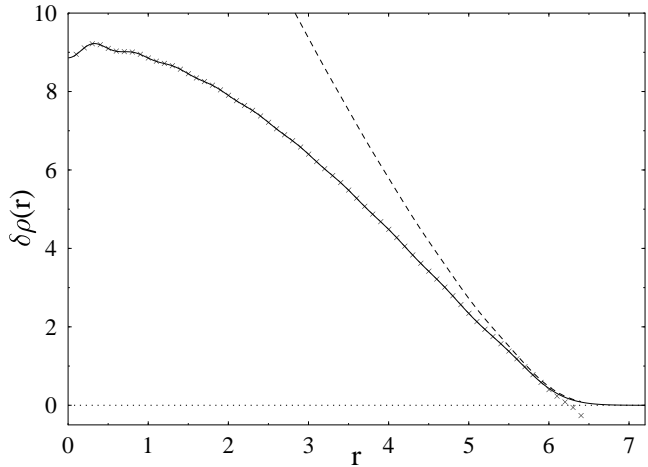}\vspace*{-0.4cm}
\caption{\label{3horhotot}
Total particle density for the $3D$ IHO with $N=3080$
particles ($M_{\text{s}}=20$) (units $\hbar=m=\omega=1$).
{\it Solid lines:} exact result. {\it Crosses:}
semiclassical result for $\delta\rho(r)$ in \eq{rho_HO}, summed up
to $k_{\text{max}}=15$, plus $\rho_{\text{ETF}}(r)$.
{\it Dashed line:} uniform linearized approximation  (\ref{rholin3}) 
from \cite{nlvt} with argument ${\widetilde z}_\lambda$ in \eq{zunirad}.
}
\end{figure}
In \fig{3horhotot} we show the total density for the $3D$ IHO
with $M_{\text{s}}=20$ filled shells. The solid line is the exact quantum
result \eq{rho}. The crosses give the semiclassical result as
the sum $\rho_{\text{ETF}}(r)+\delta\rho(r)$, where the latter is
calculated from the sum over the NPOs in \eq{rho_HO} up to
$k_{\text{max}}=15$. We see that the semiclassical result reproduces
very accurately the exact result up to $r\sim 5.9$, which is
rather close to the turning point $r_\lambda\sim 6.48$ where it
diverges. The linearized approximation is shown by the dashed
line; it approximates the exact density closely above $r\sim 5.8$.
Thus, switching from the semiclassical approximation to the
linearized one around $r\sim 5.85$ allows one to obtain a very
good approximation of the density in all points.

\subsection{Billiard systems}
\label{secfriedbil}

In billiards with reflecting walls, the above linearization
is not possible since the slope of the potential is always
infinite at the classical turning points. The amplitude of
the primitive ``+'' orbit can in such systems be regularized
by using the following uniform approximation of the Green
function for short times \cite{agam,bemo}:
\begin{eqnarray}
G_{\text{scl}}^{\text{(un)}}(E,{\bf r},{\bf r'})&=&\frac{m \pi}{i \hbar (2 \pi \hbar)^{D/2}} \sum_{\gamma}
\bigg|\frac{S}{p_{_{\parallel}}p'_{_{\parallel}}}
\det \frac{\partial{\bf p}_{\!\bot}}{\partial{\bf r'}_{\!\bot}}\bigg|^{1/2} \nonumber\\
&&\hspace{.5cm} \times~ H^{(1)}_{D/2-1}\bigg ( S/\hbar-\mu \pi/2 \bigg ),\label{green_st}
\end{eqnarray}
where $H^{(1)}_{\nu}(x)$ is the Hankel function of the first kind.
To evaluate the corresponding uniform approximation for the
particle density, we have to take the imaginary part of
\eq{green_st} and perform the integration over the energy.
This last step is not easily done analytically in general,
since $H^{(1)}_{\nu}(x)$ is not a simple oscillatory function
of the energy. In the following, we give results for the
contributions to the particle density $\rho(\bfr)$ in two
special cases. Unfortunately, we have not been able to derive
the corresponding contributions to the kinetic-energy densities.

\subsubsection{Arbitrary 2$D$ billiard}
\label{secfriedel2d}

For billiards in $D$=2 dimensions with arbitrary boundaries, the
uniform contribution to the particle density becomes (see \cite{agam}
for details):
\begin{eqnarray}
\delta \rho_+^{\text{(un)}}(d) = -\frac{p_{\lambda} J_1(2 d p_{\lambda}/\hbar )}
                 {2\pi \hbar d \sqrt{1-d/R}}\,,
\label{friedreg2d}
\end{eqnarray}
where $d$ is the distance from the boundary and $R$ its curvature
radius at the reflection point. Hereby it is assumed that $d$ is small
enough so that there is only one ``+'' orbit going to the boundary and
back to the given starting point. Note that the curvature radius $R$
is negative if the boundary is convex at the turning point.

\subsubsection{Spherical billiards in $D$ dimensions}

For spherical billiards in $D$ dimensions with radius $R$, the
energy integral over \eq{green_st} can also be performed, and the
regularized contribution of the primitive ``+'' orbit becomes
\be
\delta\rho_{+}^{\text{(un)}}(r) = -\rho_{\text{TF}}^{(D)}\,2^\nu\Gamma(\nu+1)
                           \left(\frac{R}{r}\right)^{\!\nu-1/2}
                           \frac{J_\nu(z)}{z^\nu}\,,
\label{drfriedd}
\ee
where $\rho_{\text{TF}}^{(D)}$ is the TF density given in \eq{rhotf}, and
\be
\nu=D/2\,, \qquad z=2\,(R-r)\,p_\lambda/\hbar\,.
\ee
For $D=3$, the expression \eq{drfriedd} agrees with a result derived
by Bonche \cite{bonc} using the multiple-reflection expansion
of the Green function introduced by Balian and Bloch \cite{babl}.
For $D=1$ (one-dimensional box), the result \eq{drfriedd} is also
found from the exact solution.

As mentioned above, the contribution \eq{drfriedd} is responsible for
the Friedel oscillations in the densities near the boundary $r=R$.
It is interesting to perform the spatial integral of \eq{drfriedd} over
the volume of the billiard. Using the formula (\cite{grry}, 6.561.14,
with $\mu=-\nu$)
\be
\int_0^\infty\frac{J_\nu(x)}{x^\nu}\,dx = \frac{1}{2^\nu}\,
                        \frac{\Gamma(1/2)}{\Gamma(\nu+1/2)}\,,
\ee
the integral can be done in the limit $p_\lambda\to\infty$ (i.e.,
for large particle numbers), and the asymptotically leading term
yields the following contribution to the particle number:
\be
\delta N_{\cal S} \simeq - \frac{1}{2\pi^{D/2}\hbar^{D-1}}\,
       \frac{\Gamma(D/2)}{\Gamma(D)}\,p_\lambda^{D-1}\,{\cal S}_D\,,
\label{weylsurf}
\ee
where ${\cal S}_D$ is the hypersurface of the $D$-dimensional sphere:
\be
{\cal S}_D = \frac{2\pi^{D/2}}{\Gamma(D/2)}\,R^{D-1}\,.
\ee

We note that \eq{weylsurf} corresponds precisely to the
surface term in the Weyl expansion \cite{weyl} of the
particle number $N$. The Fermi energy $\lambda_{\text{TF}}$ in
\eq{rhotf} hereby has to be replaced by the corresponding
quantity $\lambda_{\text{Weyl}}$ obtained by integrating the
Weyl-expanded density of states to the particle number $N$.

The role of the ``+'' orbit in contributing the surface
term to the Weyl expansion of the density of states has been
demonstrated by Zheng \cite{zheng} for arbitrary billiards in
$D=2$ dimensions.


\section{General results for finite fermion systems}
\label{traps}

After having presented our semiclassical theory for spatial
density oscillations and tested it in various model potentials,
we shall now discuss some of its results in the general context 
of finite fermion systems. Besides the trapped fermionic gases 
\cite{jin} mentioned already in the introduction, we have in mind 
also self-bound molecular systems with local pseudo\-potentials, 
such as clusters of alkali metals \cite{clust}, treated in the
mean-field approach of DFT with the local density approximation 
(LDA), for which the KS-potential is local \cite{dft}. 

\subsection{Local virial theorem}
\label{seclvt}

One of our central results was given in Eq.\ \eq{lvt} which
we repeat for the present discussion:
\begin{equation}
\delta \tau({\bf r})\simeq[\lambdab-V({\bf r})] \,\delta \rho({\bf r}) \,.
\label{lvtr}
\end{equation}
We call it the ``local virial theorem'' (LVT) because it connects
the oscillating parts of the kinetic and potential energy densities 
locally at any given
point $\bfr$. While the well-known virial theorem relates, both 
classically and quantum-mechanically, {\it integrated} (i.e., 
averaged) kinetic and potential energies to each other, the LVT 
in \eq{lvtr} does this {\it locally} at any point $\bfr$. We 
recall that hereby the Fermi energy $\lambdab$ of the averaged 
system is defined by Eq.\ \eq{nsmooth}. 

Since no particular assumptions need be made \cite{noteB} to derive 
\eq{lvtr} semiclassically from the basic equations \eq{drhosc} and
\eq{dtausc}, the LVT holds for arbitrary local potentials, and 
hence also for systems of interacting fermions in the mean-field 
approximation given by the DFT-LDA-KS approach. This is in itself 
a interesting basic result. It may also be of practical interest,
because it allows one to determine kinetic energy densities from the 
knowledge of particle densities that in general are more easy to
measure experimentally. We leave it as a challenge to the
condensed matter community, in particular those working with
trapped ultracold fermionic atoms, to verify the LVT experimentally.
 
Other forms of local virial theorems have been derived in
\cite{bm} from the exact quantum-mechanical densities of
isotropic harmonic oscillators in arbitrary dimensions. A
Schr\"odinger-like (integro-)differential equation for the
particle density $\rho(r)$ has also been derived in \cite{bm}.
It would lead beyond the scope of the present paper to discuss
these results and their generalization to arbitrary local
potentials based upon our semiclassical theory. This will be
done in a forthcoming publication \cite{nlvt}, where we also
give exact expressions for spatial densities in linear
potentials of which we already have made use in \sec{secsurf}.

\subsection{Extended validity of the TF kinetic-energy functional}

Presently we discuss the direct functional relation \eq{tautff} 
between the particle and kinetic energy densities obtained in the
Thomas-Fermi model. While Eq.\ \eq{tautff} is exact only when
applied to the TF expressions \eq{rhotf} and \eq{tautf} of the
(smooth) densities, we shall now show that a semiclassically 
approximate relation holds also between the oscillating exact 
densities:
\be
\tau(\bfr) \; \simeq \; \tau_{\text{TF}}[\rho(\bfr)]\,,
\label{tfofrho}
\ee
Eq.\ \eq{tfofrho} states that the TF relation \eq{tautff} holds 
approximately, for arbitrary local potentials $V(\bfr)$, also for 
the {\it exact quantum-mechanical densities including their 
quantum oscillations}. This had been observed numerically already 
earlier \cite{brfest}, but without understanding of the reason 
for its validity.

The proof of \eq{tfofrho} is actually very easy, having the LVT
\eq{lvtr} at hand. Inserting
$\rho(\bfr)=\rho_{\text{TF}}(\bfr)+\delta\rho(\bfr)$ into \eq{tautf}
and Taylor expanding around $\rho_{\text{TF}}(\bfr)$, we obtain
\be
\tau_{\text{TF}}[\rho(\bfr)] = \tau_{\text{TF}}[\rho_{\text{TF}}(\bfr)]
                      + \left.\!\frac{{\rm d}\,\tau_{\text{TF}}[\rho]}{{\rm d}\rho}
                        \right|_{\rho_{\text{TF}}(\bfr)}\!\!\!\!\!\!\delta\rho(\bfr)
                      + {\cal O}\left[(\delta\rho)^2\right]\!.
\ee
Using the obvious identity $\tau_{\text{TF}}[\rho_{\text{TF}}(\bfr)]
=\tau_{\text{TF}}(\bfr)$ and the fact that
d$\tau_{\text{TF}}[\rho_{\text{TF}}(\bfr)]/{\rm d}\rho_{\text{TF}}(\bfr)=
[\lambdab-V({\bf r})]$, we see immediately with \eq{lvtr} that, to first
order in the oscillating parts, we have indeed the relation
\be
\tau_{\text{TF}}[\rho(\bfr)] \simeq \tau_{\text{TF}}(\bfr)+\delta\tau(\bfr) = \tau(\bfr).
\label{taufct}
\ee

We stress that, although the TF expression for all three kinetic energy 
densities $\tau(\bfr)$, $\tau_1(\bfr)$ and $\xi(\bfr)$ is the same [cf.\
\eq{xitf}], the relation \eq{tfofrho} holds only for $\tau(\bfr)$. 
The reason is that the LVT also only holds for this kinetic energy
density, as discussed explicitly in the previous sections.

In \figs{chaos1} and \ref{tf2D} we present numerical tests of the
relation \eq{tfofrho} for the two-dimensional coupled quartic oscillator 
\eq{vqo}, which represents a classically chaotic system, with two
different particle numbers. An example for the three-dimensional spherical 
billiard, which is a good approximation for the self-consistent 
mean field of very large alkali metal clusters \cite{sciam}, is shown in 
\fig{tf3D}. We see that in all cases, the relation \eq{tfofrho} between the 
exact quantum-mechanical densities $\tau(\bfr)$ and $\rho(\bfr)$ is extremely
well fulfilled; only close to the classical turning points, where the
LVT \eq{lvtr} does not hold, do we see a slight deviation. Obviously,
the terms of order ${\cal O}\left[(\delta\rho)^2\right]$, neglected
in the above derivation, play practically no significant role in the
interior of the systems -- even for moderate particle numbers $N$ as
seen in \fig{tf2D} or in the examples given in Ref.\ \cite{brfest}
(and reproduced in \cite{bvz}).
\begin{figure}[h]
\includegraphics[width=1.\columnwidth,clip=true]{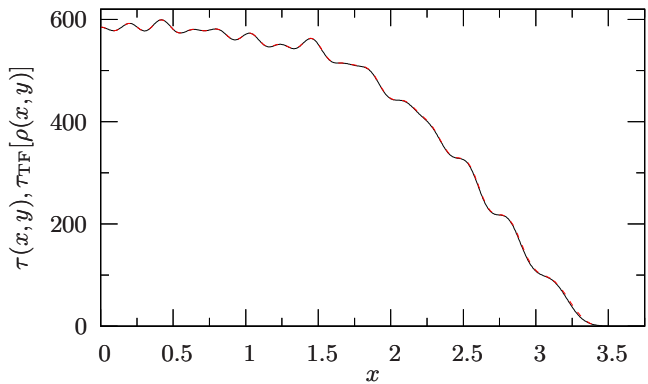}
\vspace*{-0.6cm}\caption{\label{chaos1}
(Color online) TF functional relation \eq{tfofrho} for the same system
as in \fig{chaos} ($N=632$ particles).
Cuts along the diagonal $x=y$. The solid (black) line is the l.h.s., and
the dashed (red) line is the r.h.s.\ of \eq{tfofrho}.
}
\end{figure}
\begin{figure}[h]
\includegraphics[width=1.\columnwidth,clip=true]{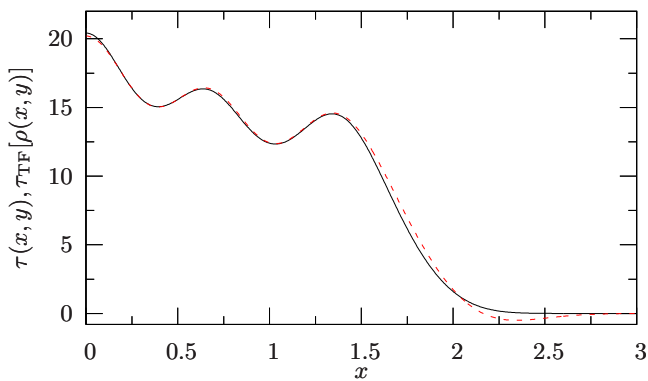}
\vspace*{-0.3cm}
\caption{\label{tf2D}
(Color online) Same as in \fig{chaos1} for $N=42$ particles.
Cuts along the line $y=x/\sqrt{3}$.
}
\end{figure}
\begin{figure}[h]
\hspace*{0.3cm}\includegraphics[width=1\columnwidth,clip=true]{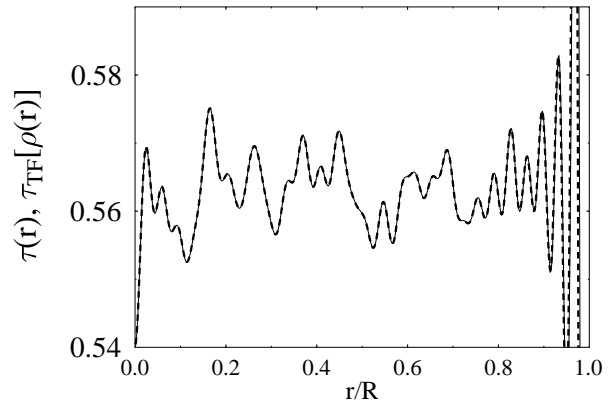}
\vspace*{-0.5cm}
\caption{\label{tf3D}
Test of the TF functional relation \eq{tfofrho} for $N=100068$ particles
in the three-dimensional spherical billiard (lines as in \fig{tf2D},
units $\hbar^2\!/2m=R=1$; both densities divided by $N^{5/3}$).
Note that in the vertical direction of the figure, only a very small
excerpt around the bulk value is displayed.
}
\end{figure}

This result might come as a surprise, since it is well
known from the ETF model that {\it for smooth densities} the gradient
corrections to the functional $\tau_{\text{TF}}[\rho]$ do play an 
important role for obtaining the correct average kinetic energy. 
(For three-dimensional systems, the first of them is 
the famous Weizs\"acker correction \cite{ws}.) Examples for this are 
given in chapter 4.4 of \cite{book}. However, if gradient corrections 
up to a given order were consistently added to \eq{tfofrho} and used 
with the {\it exact} density $\rho(\bfr)$, the agreement seen in the 
above figures would be completely spoiled.

\section{Summary and concluding remarks}
\label{secsum}

We have presented a semiclassical theory, initiated in \cite{rb}, 
for the oscillating parts of
the spatial densities in terms of closed non-periodic orbits (NPOs),
while the smooth part of the densities are given by the (extended)
Thomas-Fermi (TF) theory. Our equations \eq{drhosc} -- \eq{dtau1sc}
are the analogues of the semiclassical trace formula \eq{trf} for
the density of states in terms of periodic orbits.

For spherical systems, two kinds of oscillations in
the spatial densities can be separated, as is implied in Eqs.\
\eq{drhodec} -- \eq{dxidec}: regular, short-ranged ones
(denoted by the symbol $\delta_{\text{r}}$)
that we can attribute to the librating NPOs in the radial direction,
and irregular, long-ranged ones (denoted by $\delta_{\text{irr}}$)
that are due to non-radial NPOs and therefore only exist in $D>1$
dimensions. The simple nature of the radial NPOs leads immediately
to a number of relations between the regular parts of the
oscillations, such as Eqs.\ \eq{tautau1r}, \eq{lapeqv}, or the
universal form \eq{delrhorad} for $\delta_{\text{r}}\rho(r)$ valid
near $r=0$. It also explains that the kinetic-energy density
$\xi(r)$ defined in \eq{xi} has no rapid regular oscillations,
as implied in \eq{dxidec}, but is smooth for all  one-dimensional
systems, as well as for isotropic harmonic oscillators \cite{bm}
and linear potentials \cite{nlvt} in arbitrary $D$ dimensions, 
since these contain no non-radial NPOs.

In spherical systems, the semiclassical
expansion in terms of NPOs is expected to work best for filled
``main shells'' where the total energy has a pronounced local
minimum. This is also discussed in Ref.\ \cite{circ} on the
two-dimensional circular billiard, for which a complete classification
of all NPOs (in addition to the periodic orbits) has been made and
the semiclassical theory for the spatial density oscillations has
been studied analytically. The semiclassical approximation for the
density oscillations is, indeed, found there to work best for the
closed-shell systems with filled main shells. But even for
``mid-shells'' systems with half-filled main shells and for most
intermediate systems, the agreement of the semiclassical densities
with the quantum-mechanical ones has turned out in \cite{circ} to
be very satisfactory.

Based on the semiclassical theory, we were able to generalize
the ``local virial theorem'' (LVT) given in \eq{lvt} and \eq{lvtr}, 
which had earlier been derived from exact results for isotropic 
harmonic oscillators \cite{bm}, to arbitrary local potentials $V(\bfr)$. 
We emphasize that the LVT is valid (semiclassically) also for an 
{\it interacting $N$-fermion system} bound by the self-consistent 
Kohn-Sham potential obtained within the framework of DFT and might 
be verified experimentally in finite fermionic systems.

\begin{acknowledgments}
We are grateful to M. Guti\'errez, M. Seidl, D. Ullmo, T. Kramer,
M.V.N. Murthy, A.G. Magner and S.N. Fedotkin for helpful discussions.
After posting of the first preprint of this publication, A.G. Magner 
kindly brought Ref.\ \cite{bonc} to our attention. A.K. acknowledges
financial support by the Deutsche Forschungsgemeinschaft
(Graduierten-Kolleg 638).
\end{acknowledgments}



\appendix*

\section{Inclusion of finite temperature in the semiclassical theory}
\label{appcor}

In this Appendix we give a short sketch of how to include
finite-temperatures in the semiclassical formalism. Extensions
of semiclassical trace
formulae to finite temperatures have been used already long
ago in the context of nuclear physics \cite{mako} and more
recently in mesoscopic physics \cite{ruj}. We shall present
here a derivation by means of a suitable folding function,
which has proved useful also in the corresponding microscopic
theory \cite{bq}.

For a grand-canonical ensemble of fermions embedded in a
heat bath with fixed temperature, the variational energy is
the so-called grand potential $\Omega$ defined by \cite{foot2}
\be
\Omega = \langle {\hat H} \rangle - TS - \lambda \langle {\hat N} \rangle\,,
\label{omega}
\ee
where ${\hat H}$ and ${\hat N}$ are the Hamilton and particle
number operators, respectively, $T$ is the temperature in energy
units (i.e., we put the Boltzmann constant $k_B$ equal to unity),
$S$ is the entropy, and $\lambda$ the chemical potential.
Note that both energy and particle number are conserved only
on the average. For non-interacting particles, we can write
the Helmholtz free energy $F$ as
\be
F = \langle {\hat H} \rangle - TS = 2\sum_n E_n \nu_n - TS\,,
\label{free}
\ee
where $E_n$ is the energy spectrum of ${\hat H}$ and $\nu_n$
are the Fermi occupation numbers
\begin{equation}
\nu_n = \frac{1}{1+\exp{\left(\frac{E_n - \lambda}{T}\right)}}\,,
\label{nuocc}
\end{equation}
and the entropy is given by
\begin{equation}
S=-2\sum_n \,[\nu_n \log\nu_n + (1-\nu_n) \log (1-\nu_n)] \,.
\label{Sent}
\end{equation}
The chemical potential $\lambda$ is determined by fixing the
average particle number
\be
N = \langle {\hat N} \rangle = 2\sum_n \nu_n\, .
\label{avnum}
\ee

It can be shown \cite{bq} that the above quantities
$N$, $F$ and $S$ may be expressed in terms of a convoluted {\it
finite-temperature level density} $g_T(E)$ as
\begin{equation}
F=2\int_{-\infty}^\lambda E\,g_T(E)\,\d E\,.
\label{gtrel}
\end{equation}
The function $g_T(E)$ is defined by a convolution of the
``cold'' ($T=0$) density of states \eq{dos}
\begin{equation}
g_T(E)=\int_{-\infty}^\infty g(E')\,f_T(E-E') \, \d E'
      =\sum_n f_T(E-E_n)\,,
\label{gT}
\end{equation}
whereby the folding function $f_T(E)$ is given as
\be
f_T(E) = \frac{1}{4T\,\Cosh^2(E/2T)}\,.
\label{fT}
\ee
Note that all sums in \eq{free} -- \eq{fT} run over the
complete (infinite) spectrum of the Hamiltonian ${\hat H}$.
It is now easily seen that
\be
N = 2 \int_{-\infty}^\lambda g_T(E)\,\d E\,.
\ee
To show that the integral \eq{gtrel} gives, indeed, the correct
free energy \eq{free} including the ``heat energy'' $-TS$ needs some
algebraic manipulations. From $F$, the entropy $S$ can always be
gained by the canonical relation
\be
S=-\papa{F}{T}\,.
\label{canonent}
\ee

The same convolution can now be applied also to the semiclassical
trace formula \eq{trf} for the oscillating part of the density
of states which we re-write as
\be
\delta g(E) \simeq \text{Re} \sum_{\text{PO}} {\cal A}_{\text{PO}}(E)\,
                             e^{\frac{i}{\hbar}S_{\text{PO}}(E)-i\sigma_{\text{PO}}}.
\label{trfc}
\ee
The oscillating part $\delta g_T(E)$ of the finite-temperature level density is
obtained by the convolution of \eq{trfc} with the function $f_T(E)$
as in \eq{gT}. In the spirit of the stationary-phase approximation,
we take the slowly varying amplitude ${\cal A}_{\text{PO}}(E)$ outside of
the integration and approximate the action in the phase by
\be
S_{\text{PO}}(E') \simeq S_{\text{PO}}(E) + (E'-E)\,T_{\text{PO}}(E)\,,
\ee
so that the result becomes a modified trace formula
\be
\delta g_T(E) \simeq \text{Re} \sum_{\text{PO}} {\cal A}_{\text{PO}}(E)\,
                     {\tilde f}_T(\Tau_{\text{PO}}(E))\,
                     e^{\frac{i}{\hbar}S_{\text{PO}}(E)-i\sigma_{\text{PO}}},
\label{trft}
\ee
where
\be
\Tau_{\text{PO}}(E) = T_{\text{PO}}(E)/\hbar\,
\ee
and the temperature modulation factor ${\tilde f}_T$ is given
by the Fourier transform of the convolution function $f_T$:
\be
{\tilde f}_T(\Tau) = \int_{-\infty}^\infty f_T(\omega)\,e^{i\Tau\omega}\, \d \omega\,.
\ee
The Fourier transform of the function \eq{fT} is known, so that
\be
{\tilde f}_T(\Tau) = \frac{\pi T\Tau}{\Sinh(\pi T\Tau)}\,.
\label{modT}
\ee
The ``hot'' trace formula \eq{trft} with the modulation factor
\eq{modT} has been obtained in \cite{ruj,mako}.

For the spatial densities, we can proceed exactly in the same way.
For the particle density, e.g., the microscopic expression \eq{rho}
is replaced by
\be
\rho_T(\bfr) = 2 \sum_n |\phi_n(\bfr)|^2\nu_n\,,
\label{rhoT}
\ee
where the sum again runs over the complete spectrum. Starting
from the semiclassical expression \eq{drhosc} for $\delta\rho(r)$
at $T=0$, we rewrite it as
\be
\delta\rho_0(\lambdab,\bfr) \simeq \text{Re} \sum_{\text{NPO}} {\cal A}_{\text{NPO}}(\lambdab,\bfr)\,
                                   e^{i\Phi(\lambdab,\bfr)},
\label{drhoscfoll}
\ee
where the amplitude $\cal{A}_{\text{NPO}}$ collects all the prefactors of
the phase in \eq{drhosc}. The finite-$T$ expression is given by
the convolution integral
\be
\delta\rho_T(\lambdab,\bfr) \simeq \int_{-\infty}^{\lambdab} \delta\rho_0(\lambdab-E,\bfr)
                                   f_T(E)\, \d E\,.
\ee
Expanding the phase under the integral as above, we arrive at
\be
\delta\rho_T(\lambdab,\bfr) \simeq \text{Re} \sum_{\text{NPO}} {\cal A}_{\text{NPO}}(\lambdab,\bfr)\,
                            {\tilde f}_T(\Tau_{\text{NPO}}(\lambdab,\bfr))\,e^{i\Phi(\lambdab,\bfr)}\,,
\label{drhoscT}
\ee
where $\Tau_{\text{NPO}}=T_{\text{NPO}}(\lambdab,\bfr)/\hbar$ is the period of the NPO
in units of $\hbar$. The corresponding expressions for the other spatial
densities are obvious.

For the smooth parts of the densities, we recall that the (E)TF theory
at $T>0$ is well known and refer to chapter 4.4.3 of \cite{book} for
the main results and relevant literature.

Other types of correlations can be included in the semiclassical
theory in the same way, as soon as a suitable folding $f_{\text{corr}}(E)$
function corresponding to $f_T(E)$ in \eq{fT} and its Fourier transform
are known (see. e.g., Ref.\ \cite{kaz09}).

\vfill


\begin{table*}[hb]
\caption{\label{table1}
Contributions of different types of non-periodic orbits to the spatial
densities in a rectangle billiard with sides $Q_x$ and $Q_y$.
The first row gives the position of the images of $P$ with $(k_x,k_y)
\in \mathbb{Z}^2$. The second row gives the length of
the orbit and the third row the angle $\theta$ between the
initial and final momentum.
The fourth and fifth rows give the contributions to $\delta
\rho$ and $\delta \tau_1$, respectively.
}
\begin{tabular}{llllllllll}
\hline
\hline
 & & &&&&& \\
\bf{Type of orbits}&&&a&&&b&&&c\\
& & &&&&& \\
\hline
& & &&&&& \\
\bf{Image points of $P(x,y)$ }
&&&$(2 k_x Q_x+x,2 k_y Q_y-y)$
&&&$(2 k_x Q_x-x,2 k_y Q_y+y)$
&&&$(2 k_x Q_x-x,2 k_y Q_y-y)$\\
 &&& & &&& \\
\hline
 & &&& &&& \\
\bf{Orbit length}
&&&$\displaystyle L(k_x Q_x,k_y Q_y-y)$
&&&$\displaystyle L(k_x Q_x-x,k_y Q_y)$
&&&$\displaystyle L(k_x Q_x-x,k_y Q_y-y)$\\
&&&& & & & \\
\hline
&&&& & & & \\
\bf{$\theta$}
&&&$\theta_{\text{a}}= -2 \arctan \bigg ( \frac{k_y Q_y-y}{k_x Q_x} \bigg)$
&&&$\theta_{\text{b}}= 2 \arctan \bigg ( \frac{k_y Q_y}{k_x Q_x-x} \bigg)$
&&&$\theta_{\text{c}}= \pi$\\
&&&& & & & \\
\hline
&&&& & & & \\
\bf{Contribution to $\delta \rho$}
&&&$\delta \rho_{\text{a}} = f(k_x Q_x,k_y Q_y-y,1)$
&&&$\delta \rho_{\text{b}} = f(k_x Q_x-x,k_y Q_y,1)$
&&&$\delta \rho_{\text{c}} = f(k_x Q_x-x,k_y Q_y-y,0)$\\
&&&& & & & \\
\hline
&&&& & & & \\
\bf{Contribution to $\delta \tau_1$}
&&&$\delta \tau_{1{\text{a}}}= \lambdab \cos(\theta_{\text{a}}) \delta \rho_{\text{a}}$
&&&$\delta \tau_{1{\text{b}}}= \lambdab \cos(\theta_{\text{b}}) \delta \rho_{\text{b}}$
&&&$\delta \tau_{1{\text{c}}}= \lambdab \cos(\theta_{\text{c}}) \delta \rho_{\text{c}}$\\
&&&& & & & \\
\hline
\hline
\end{tabular}
\end{table*}


\begin{thebibliography}{31}
\bibitem{foot} Note that in the standard literature on DFT, $\tau(\bfr)$
               sometimes denotes what we here call $\tau_1(\bfr)$. See
               also Ref.\ \cite{dft}, chapter 5.5, for a discussion
               and further literature on the various forms of the
               kinetic-energy density.
\bibitem{notebil} In billiard systems with hard-wall reflection, there
                  exists no gradient expansion of the potential and
                  therefore the smooth parts of the densities are
                  given by their TF values.
\bibitem{conjp} Conjugate points are those in which a
                fan of slightly perturbed orbits, obtained by small
                changes of the initial momentum (or position) of a
                given orbit, focuses again on the unperturbed orbit, 
                see \cite{gubu}.
\bibitem{noteB}  We point out that the LVT holds also at critical points $\bfr$
                 connected with symmetry breaking and bifurcations, where the 
                 semiclassical amplitudes in \eq{drhosc} -- \eq{dtau1sc}
                 have to be regularized by uniform approximations (cf.\
                 \sec{secregul}; see also Ref.\ \cite{circ} for the 
                 circular billiard).
\bibitem{foot2}  Note that in this Appendix, $T$ and $S$ without subscript denote
                 temperature and entropy, respectively, while the same symbols
                 with subscripts ``PO'' or ``NPO'' denote the periods and actions
                 of the classical orbits, as used everywhere else in this paper.
\end{thebibliography}
\end{document}